\documentclass[authoryear,5pt]{elsarticle}

\usepackage{lineno,hyperref}

\usepackage{fullpage}

\usepackage{amsmath}
\usepackage{amsfonts}
\usepackage{caption}
\usepackage{subcaption}
\usepackage{soul}

\usepackage{natbib}
\usepackage{algorithm}% http://ctan.org/pkg/algorithms
\usepackage{algpseudocode}%
\usepackage{booktabs}
\usepackage{epstopdf}

\usepackage{physics}
\usepackage[abs]{overpic}
\usepackage{tikz}
\usepackage{wasysym}
\usepackage{relsize}
\usepackage{pifont}
\usepackage{xcolor}
\usepackage{tabularx}
\usepackage{amsmath}
\usepackage{amssymb}
\usepackage{siunitx}
\usepackage{natbib}
\usepackage{amsmath}

\newcommand{\rout}[1]

\setcitestyle{authoryear,open={(},close={)}}

\usepackage{graphicx}
\usepackage{float}
\usepackage[english]{babel}
\usepackage{hyperref}
\usepackage{amssymb}
\usepackage[normalem]{ulem}
\usepackage{amsmath}
\usepackage{amsthm}
\usepackage{amsfonts}
\usepackage{enumerate}
\usepackage{mathtools}
\usepackage{lineno}
\usepackage{bm}

\modulolinenumbers[5]

\journal{Elsevier}

\graphicspath{{figure/}}

\usepackage{amssymb}

\errorcontextlines\maxdimen

\makeatletter
\newcommand*{\algrule}[1][\algorithmicindent]{\makebox[#1][l]{\hspace*{.5em}\vrule height .75\baselineskip depth .25\baselineskip}}%

\newcount\ALG@printindent@tempcnta
\def\ALG@printindent{%
    \ifnum \theALG@nested>0% is there anything to print
        \ifx\ALG@text\ALG@x@notext% is this an end group without any text?
            \addvspace{-3pt}% FUDGE for cases where no text is shown, to make the rules line up
        \else
            \unskip
            \ALG@printindent@tempcnta=1
            \loop
                \algrule[\csname ALG@ind@\the\ALG@printindent@tempcnta\endcsname]%
                \advance \ALG@printindent@tempcnta 1
            \ifnum \ALG@printindent@tempcnta<\numexpr\theALG@nested+1\relax% can't do <=, so add one to RHS and use < instead
            \repeat
        \fi
    \fi
    }%
\usepackage{etoolbox}
\patchcmd{\ALG@doentity}{\noindent\hskip\ALG@tlm}{\ALG@printindent}{}{\errmessage{failed to patch}}
\makeatother
\begin{document}

\begin{frontmatter}

\title{Real-time simulation enabled navigation control of magnetic soft \\ continuum robots in confined lumens}

\author{Dezhong Tong$^{1,2,\dagger}$,  Zhuonan Hao$^{3,\dagger}$, Jiyu Li$^{1,\dagger}$, Boxi Sun$^{1}$, Mingchao Liu$^{4,\star}$, \\ Liu Wang$^{1,\star}$, Weicheng Huang$^{5,\star}$}

\address{
$^{1}$CAS Key Laboratory of Mechanical Behavior and Design of Materials, Department of Modern Mechanics, \\ University of Science and Technology of
China, Hefei 230026, PR China\\
$^{2}$Department of Material Science and Engineering, University of Michigan, Ann Arbor, \\ Ann Arbor, Michigan, 48105, USA \\
$^{3}$Department of Mechanical and Aerospace Engineering, University of California, Los Angeles,  \\
Los Angeles, California 90095, United States\\
$^{4}$Department of Mechanical Engineering, University of Birmingham, Birmingham B15 2TT, UK\\
$^{5}$School of Engineering, Newcastle University, Newcastle upon Tyne, NE1 7RU, UK\\
$^{\dagger}$These authors contributed equally to this work. \\
$^{\star}$Corresponding authors:\\ 
\ m.liu.2@bham.ac.uk (M.L.)  \\ wangliu05@outlook.com (L.W.) \\ weicheng.huang@ncl.ac.uk (W.H.).
}

\begin{abstract}

Magnetic soft continuum robots (MSCRs) have emerged as a promising technology for minimally invasive interventions, offering enhanced dexterity and remote-controlled navigation in confined lumens.
Unlike conventional guidewires with pre-shaped tips, MSCRs feature a magnetic tip that actively bends under applied magnetic fields.
Despite extensive studies in modeling and simulation, achieving real-time navigation control of MSCRs in confined lumens remains a significant challenge.
The primary reasons are due to robot-lumen contact interactions and computational limitations in modeling MSCR nonlinear behavior under magnetic actuation.
Existing approaches, such as Finite Element Method (FEM) simulations and energy-minimization techniques, suffer from high computational costs and oversimplified contact interactions, making them impractical for real-world applications.
In this work, we develop a real-time simulation and navigation control framework that integrates hard-magnetic elastic rod theory, formulated within the Discrete Differential Geometry (DDG) framework, with an order-reduced contact handling strategy.
Our approach captures large deformations and complex interactions while maintaining computational efficiency.
Next, the navigation control problem is formulated as an inverse design task, where optimal magnetic fields are computed in real time by minimizing the constrained forces and enhancing navigation accuracy.
We validate the proposed framework through comprehensive numerical simulations and experimental studies, demonstrating its robustness, efficiency, and accuracy.
The results show that our method significantly reduces computational costs while maintaining high-fidelity modeling, making it feasible for real-time deployment in clinical settings.
Our work addresses key limitations in MSCR navigation control, paving the way for safer and more reliable clinical translation of MSCR technology for interventional surgeries. 

\end{abstract}


\end{frontmatter}

\section{Introduction}	

% logic: 
% 1. Medical treatment and interventional procedures needs the magnetic wires
% 2. To facilitate the application of magnetic wires, modelling and analysis of MSCRs have been extensively studied.
% 2.1 from 2D to 3D, theoretical (analytical solutions)
% 2.2

%\WH{Add more citations for the first two and last four paragraphs.}

Interventional surgeries are a standard treatment for vascular and gastrointestinal diseases, offering reduced recovery times and fewer complications compared to traditional open surgeries \citep{muller1992peripheral}.
In endovascular interventions such as thrombectomy and aneurysm embolization, a mechanical guidewire with a pre-shaped tip is inserted through a small incision into the femoral artery and manually navigated under fluoroscopic guidance to reach the target site \citep{munich2019overview}. 
However, the pre-shaped tip of mechanical guidewires poses 
significant challenges in navigating tortuous or branching confined lumens \citep{goyal2016endovascular,appireddy2016endovascular}.
The reliance on trial-and-error maneuvering increases procedure time, risks vessel wall injury, and exposes both the patient and surgeon to prolonged fluoroscopic radiation \citep{zhao2022remote,duan2023technical}.

\begin{figure}[ht]
\centering
\includegraphics[width=1.0\columnwidth]{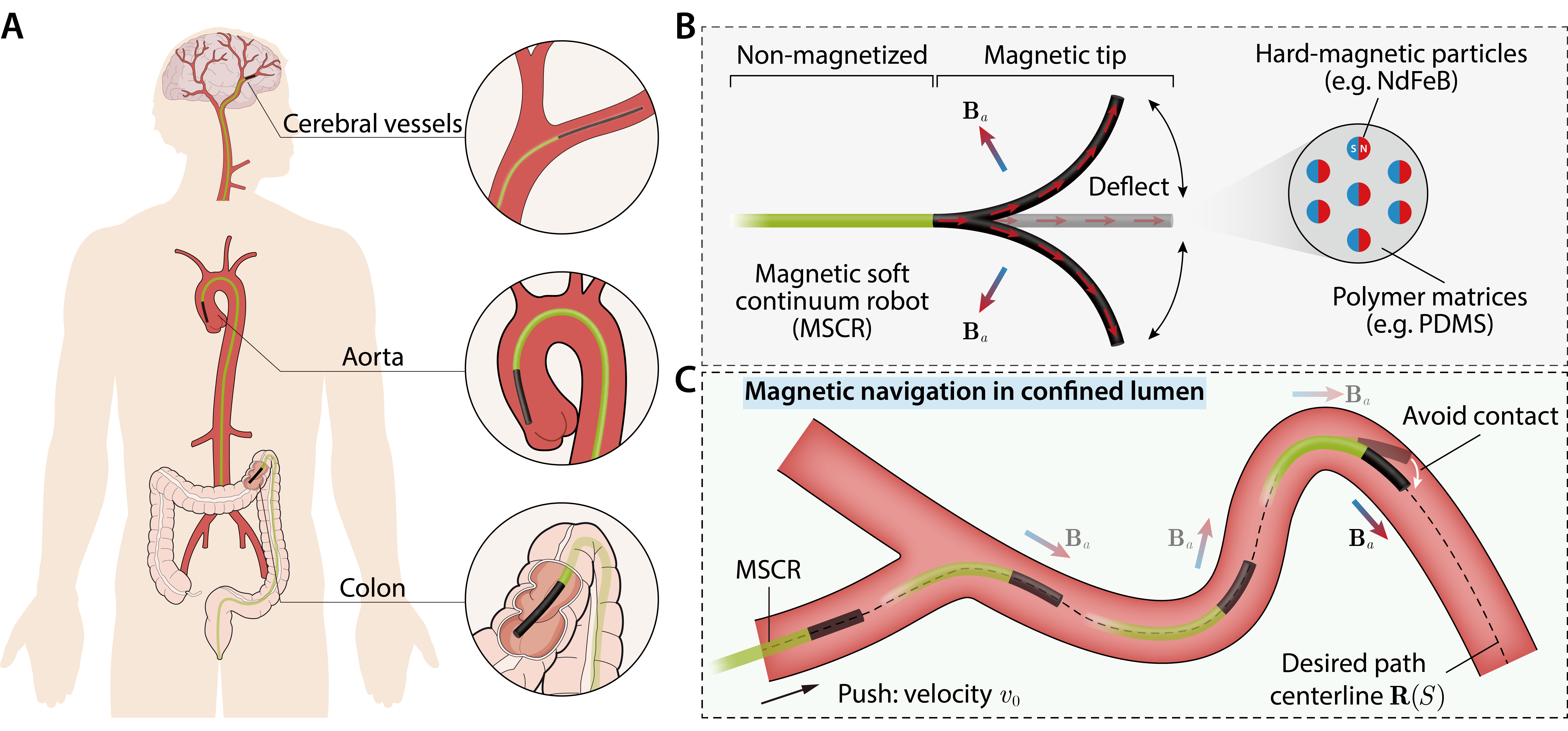}
\caption{Schematic of a magnetic soft continuum robot (MSCR) navigating through the confined lumen within the biological systems. (A) Anatomical illustration of luminal structures, including cerebral vessels, the aorta, and the colon. The MSCR demonstrates adaptability for operation within diverse pathways. (B) Composition of the MSCR. The magnetic tip (black), fabricated by embedding hard-magnetic particles (e.g. NdFeB) within polymer matrices (e.g. PDMS), deflects in response to an actuation magnetic field $\mathbf{B}_{a}$. (C) The MSCR is pushed at the proximal end with velocity $v_0$. By controlling the actuation magnetic fields $\mathbf{B}_{a}$, the MSCR is steered along the desired path (centerline $\mathbf{R}(S)$) through the confined lumen without the tip contact with the lumen wall.  }
\label{fig:overview}
\end{figure}

To overcome these challenges, magnetic soft continuum robots (MSCRs) have recently emerged as a promising alternative. Distinct from the pre-shaped tip, MSCRs feature a magnetically bendable tip that enhances its remotely-controlled navigation in confined lumens (as shown in Fig.~\ref{fig:overview}(A)) \citep{wu2020multifunctional,kim2022magnetic,yang2023morphing}.
Constructed from soft polymers embedded with hard-magnetic particles such as neodymium iron boron (NdFeB), the MSCR's tip actively bends in response to actuation magnetic fields (Fig.~\ref{fig:overview}(B)) \citep{kim2019ferromagnetic}. This capability enhances maneuverability in complex and confined anatomical pathways, including cerebral vessels \citep{kim2022telerobotic}, the aorta  \citep{wang2022magnetic}, and the colon \citep{martin2020enabling}, surpassing traditional guidewires.
By eliminating the need for repeated manual adjustments, MSCRs significantly reduce both procedure time and radiation exposure for patients and surgeons \citep{hwang2020review,gunduz2021robotic}. 

The promising potential of MSCRs has driven extensive research into their magneto-mechanical behavior through theoretical modeling and numerical simulation.
Early studies by \cite{zhao2019mechanics} introduced a hyperelastic constitutive model to describe the interactions between dispersed magnetic particles and applied magnetic fields using magnetic Cauchy stresses. 
\cite{wang2020hard} extended this by developing a hard-magnetic elastica model to describe the large deflections under uniform magnetic fields, providing analytical solutions for MSCR deformations. 
Further refinements have included extensible rod models that accounts for stretching deformations \citep{chen2020complex,chen2020mechanics,chen2020theoretical} and a beam-based framework incorporating three-dimensional (3D) Finite Element Method (FEM) simulation \citep{yan2022comprehensive}. 
In addition, more complex approaches have incorporated nonlinear large deformation analyses, using Kirchhoff-like rod models \citep{sano2022kirchhoff}, Euler-angle-based formulations \citep{chen2021three}, and models that consider deformable cross-sections \citep{li2023mechanics}.

Beyond analytical modeling, numerical methods such as finite difference approximations \citep{wang2021evolutionary}, magnetic lattice models \citep{ye2021magttice}, mesh-free models \citep{liu2023meshfree}, reduced-order FEM model \citep{moezi2024development} have been widely employed to solve complex MSCR deformation problems.
In addition, a total Lagrangian (TL) finite element formulation is constructed based on micropolar theory to capture the asymmetric stress response and size-dependent behavior of soft hard-magnetic beams~\citep{dadgar2024three}.
At the microstructural level, \cite{garcia2021microstructural} incorporated dipole-dipole interactions of embedded magnetic particles, while \cite{stewart2023magneto} developed finite deformation models for hard-magnetic viscoelastic materials.
Collectively, these studies have established a solid theoretical foundation for understanding the magneto-mechanical behavior of MSCRs in static and uniform magnetic fields.

Recent advances have expanded MSCR modeling to include dynamic behavior and nonuniform magnetic fields.
For instance,  \cite{huang2023modeling, huang2023discrete} formulated a discrete magneto-elastic rod model based on Discrete Elastic Rods (DER) algorithm to analyze the dynamics of MSCR under magnetic actuation.
High-order dynamic modeling using absolute nodal coordinate formulation has also been explored by \cite{wang2024dynamic}.
\cite{li2024modeling} used energy minimization method to describe large deformations in nonuniform magnetic environments. \cite{yao2023adaptive} integrated deep reinforcement learning into Cosserat rod-based simulations to develop dynamic control strategies.

Despite these advances, precise navigation control of MSCRs under magnetic actuation still faces significant challenges \citep{yang2023magnetically}.
A key challenge arises from the complex contact interactions between the MSCR’s tip and the surrounding lumen, such as blood vessel walls \citep{khoshnam2015modeling,wang2022magnetic}. These interactions are particularly critical in confined lumens where unintended contact can alter MSCR’s trajectory and compromise navigation accuracy. More critically, excessive contact forces increase the risk of lumen injury or rupture, leading to severe complications \citep{wang2015study}. However, existing models oversimplify or neglect these contact interactions, limiting their ability to provide navigation control in real-world scenarios \citep{li2024modeling1}.

Another challenge lies in the computational demands of modeling the MSCR's dynamic behavior in confined lumens. While FEM and lattice-based models have been widely used to analyze large deformations of MSCR, their high computational cost make them impractical for real-time applications \citep{zhao2019mechanics,ye2021magttice, dadgar2022finite}. Similarly, energy minimization approaches struggle with efficiently updating contact conditions and adjusting magnetic fields in real-time, limiting their applicability in dynamic and confined environments \citep{li2024model}. These computational limitations pose a significant barrier to real-time navigation control, where the ability to rapidly predict MSCR deformations and adjust magnetic fields is critical for safe and effective operation in confined lumens.  (Fig.~\ref{fig:overview}(C)).

In this work, we address these challenges by developing a real-time simulation and navigation control framework for MSCRs in confined lumens. Our approach integrates hard-magnetic elastic rod theory with penalty energy-based contact handling, enabling computationally efficient and accurate modeling of MSCR deformations under real-time magnetic actuation. 
As shown in Fig.~\ref{fig:overview}(C), the MSCR will be pushed into the lumen without any external twisting applied on the pushing end.
By dynamically adjusting an external magnetic field, the deflection of the MSCR tip can be controlled during navigation.
To ensure safe and effective navigation, we formulate a mechanics model-based control framework that compute the optimal magnetic field in real-time based on MSCR’s geometric and material properties, as well as the surrounding lumen constraints. Consequently, the MSCR can autonomously adjust its trajectory to maintain alignment with the lumen's centerline, minimizing contact forces and reducing the risk of vessel wall injury.
We validate the performance of our model-based control framework through numerical simulations and experimental studies across various lumen geometries. 
The results demonstrate that our method achieves robust, efficient, and high-fidelity real-time performance, significantly improving MSCR navigation capability in complex environments.

This article is organized as follows. 
Section~\ref{sec:theory} presents the analytical derivation of the governing equations for a hard-magnetic rod navigating within a confined lumen. In Section~\ref{sec:simulation}, we develop a numerical framework that integrates nonlinear elasticity, magnetic actuation, and non-penetration contact handling to solve these equations. Section~\ref{sec:control} details how the model-based control problem is reformulated as an inverse design problem in mechanics. Section~\ref{sec:results} provides a comprehensive numerical study of the proposed control method, supplemented by experimental validation. Finally, we conclude in Section~\ref{sec:discussion}, discussing potential applications of the developed framework for MSCR control in medical interventions.

\section{Theoretical model}
\label{sec:theory} 

In this section, we derive the governing equations for the geometrically nonlinear deformations of a MSCR navigating through a confined lumen.
The MSCR is modeled as a 3D hard-magnetic elastic rod, while the confined lumen is represented as a curved tube.
We begin by formulating the mechanical model based on the classical Kirchhoff equations for an one-dimensional (1D) slender structure \citep{audoly2000elasticity}, then incorporate external magnetic forces and contact interactions with the lumen walls.

% In this section, we derive the governing equations for the geometrically nonlinear deformations of a hard magnetic rod moving in a constrained tube.
% %
% We first review the classical Kirchhoff equation for a 1D slender structure, next, we consider the external magnetic action as well as the contact with the cylindrical tube.

\begin{figure}[h]
\centering
\includegraphics[width=1\columnwidth]{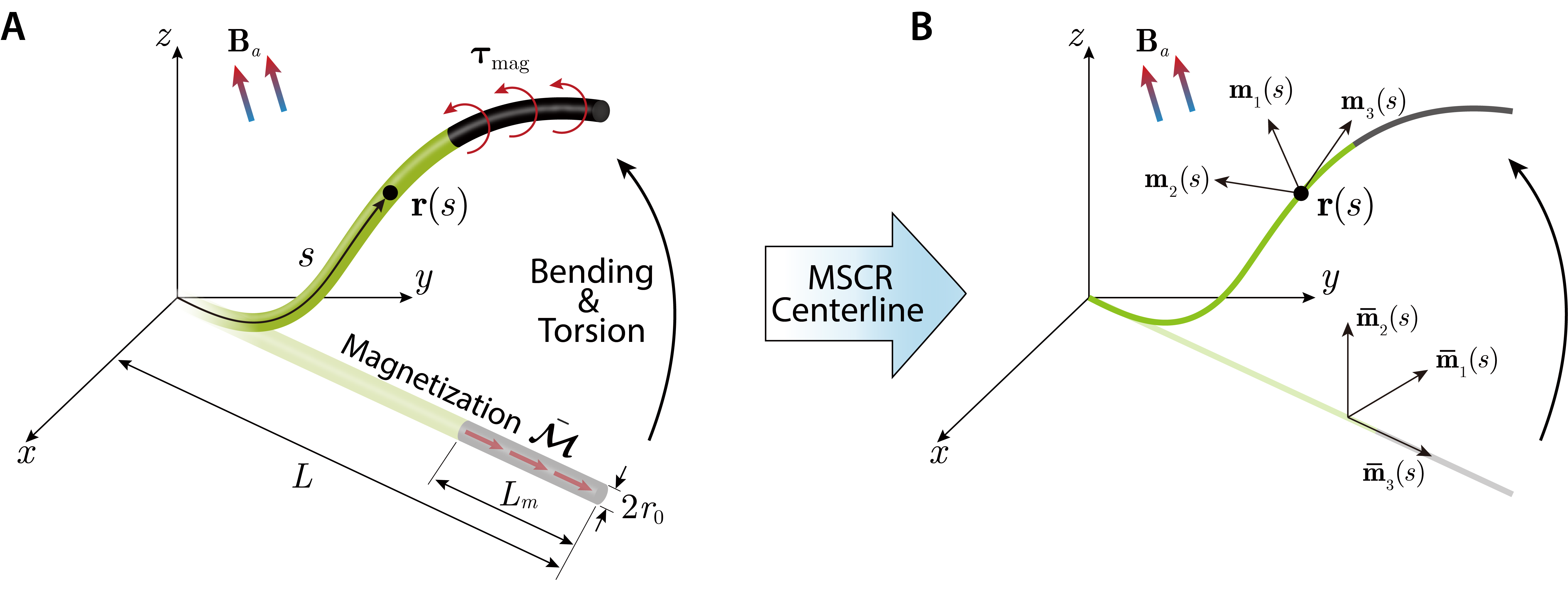}
\caption{Simplified representation of the magnetic soft continuum robot (MSCR), characterized by a total length $L$, magnetic tip length $L_{m}$, and cross-sectional radius $r_0$. The centerline is modeled as a magneto-elastic rod (denoted as $s$), with a primary focus on its bending, stretching, and twisting behaviors. }
\label{fig:schematic}
\end{figure}

\paragraph{Kinematics} 

As shown in Fig. \ref{fig:schematic}(A), a MSCR is considered as a slender rod of length $L$ and radius $r_{0}$, characterized by the second moment of area $ I = \pi r_{0}^4 / 4 $, the second polar moment of area $ J = \pi r_{0}^4 / 2 $, and the cross-sectional area $ A = \pi r_{0}^2 $. The rod is made of isotropic and linearly elastic material with Young's modulus $E$, shear modulus $G$, and density $\rho$.
The length of the magnetic tip is denoted as $L_{m}$.
Due to its slenderness, i.e., $L \gg r_{0}$, the rod can be represented by a 1D space curve parameterized by its longitudinal coordinate $s$ (Fig. \ref{fig:schematic}(B)). 
Its configuration can be described the centerline $\mathbf{r}(s)$ and an accompanying material frame $\{ \mathbf{m}_{1}(s), \mathbf{m}_{2}(s), \mathbf{m}_{3}(s) \}$, where
\begin{equation}
\mathbf{m}_{3} (s) = \frac {\mathbf{r}' (s)} { || \mathbf{r} (s) ||} ,
\end{equation}
is the tangential director, and the other two directors are orthonormal to the tangent, i.e., 
\begin{equation}
\begin{aligned}
\mathbf{m}_{1}(s) &\cdot \mathbf{m}_{2}(s) = 0 \\
\mathbf{m}_{1}(s) &\times \mathbf{m}_{2}(s) = \mathbf{m}_{3}(s).
\end{aligned}
\end{equation}
Hereafter, we use $()'$ to indicate the gradient concerning space $s$, use $\dot{()}$ to indicate the derivative with respect to time $t$, and a bar on top $\bar{()}$ represents the evaluation in the undeformed configuration, e.g., $\bar{\mathbf{r}}(s)$ is the centerline of the reference configuration and $\{ \bar{\mathbf{m}}_{1}(s), \bar{\mathbf{m}}_{2}(s), \bar{\mathbf{m}}_{3}(s) \}$ is the material frame of the reference configuration.
%
%\fix{what is the reference configuration?} \WH{reference configuration is the undeformed configuration}
%
The rod centerline is assumed to be inextensible, i.e., 
\begin{equation}
\forall s ,\;
\varepsilon (s) = 0,
\end{equation}
where $\varepsilon (s) $ is the uniaxial strain of the centerline,
\begin{equation}
\varepsilon(s) = || \mathbf{r}'(s) || - 1.
\label{eq:continousStretching}
\end{equation}
The rotation gradient of the centerline is formulated as
\begin{equation}
\begin{aligned}
\mathbf{m}'_{1} (s) &= \bm{\omega}(s) \times \mathbf{m}_{1}(s), \\
\mathbf{m}'_{2} (s) &= \bm{\omega}(s) \times \mathbf{m}_{2}(s), \\
\mathbf{m}'_{3} (s) &= \bm{\omega}(s) \times \mathbf{m}_{3}(s),
\end{aligned}
\end{equation}
where $\bm{\omega}$ is the so-called Darboux vector,
\begin{equation}
\bm{\omega} (s) = \kappa_{1}(s) \mathbf{m}_{1}(s) + \kappa_{2}(s) \mathbf{m}_{2}(s) + \kappa_{3}(s) \mathbf{m}_{3}(s).
\label{eq:continousCurvatures}
\end{equation}
Here, $\kappa_{1}$ and $\kappa_{2}$ represent the bending curvatures, while $\kappa_{3}$ is the twisting curvature.
Overall, the macroscopic strains for the 1D rod are the sum of elastic stretching strain in Eq.(\ref{eq:continousStretching}) and elastic curvatures in Eq.(\ref{eq:continousCurvatures}).

\paragraph{Total potential energy} A linear constitutive law between the curvatures and the elastic energy is used, 
\begin{equation}
\mathcal{E}_{\mathrm{ela}} = \int_{0}^{L} \left[ \frac {1} {2} EI \kappa_{1}^2(s) +  \frac {1} {2} EI \kappa_{2}^2(s) + \frac {1} {2} GJ \kappa_{3}^2(s) \right] ds.
\label{eq:ConstitutiveEquations}
\end{equation}
where $EI$ is the bending stiffness, and $GJ$ is the torsional stiffness. Also, here we assume the initial curvatures of the rod is zero, i.e., 
\begin{equation}
\bar{\kappa}_i(s) = 0, \; \mathrm{with } \; i \in \{ 1,2,3 \}.
\end{equation}
The 1D reduced magnetic energy functional is given by \citep{sano2022kirchhoff, huang2023modeling, huang2023discrete}
\begin{equation}
\mathcal{E}_{\mathrm{mag}} = - \int_{0}^{L} \left[ \bm{\mathcal{M}}(s) \cdot \mathbf{B}_{a}(s) \right] ds,
\end{equation}
where $\mathbf{B}_{a}$ is the actuation magnetic field, and $\bm{\mathcal{M}}$ is the magnetization density per unit length of the rod in the deformed configuration, which can be formulated as,
\begin{equation}
\bm{\mathcal{M}} (s) = \left[ \mathbf{m}_{1}(s) \otimes \bar{\mathbf{m}}_{1}(s) + \mathbf{m}_{2}(s) \otimes \bar{\mathbf{m}}_{2}(s) + \mathbf{m}_{3}(s) \otimes \bar{\mathbf{m}}_{3}(s) \right] \cdot \bar{\bm{{\mathcal{M}}}} (s),
\end{equation}
where $\bar{\bm{{\mathcal{M}}}}$ is the magnetization density per unit length in the undeformed configuration.

\paragraph{Confinement within a lumen} 
With a longitudinal parameter $S$, the lumen can be characterized by its centerline $\mathbf{R}(S)$ and inner radius $R(S)$, forming a 3D space $\mathcal{S}$,
\begin{equation}
\mathcal{S} = \{\mathbf P \in \mathcal{R}^3 : \lVert \mathbf P - \mathbf R(S)\rVert \leq R(S), S \in [S_\textrm{min}, S_\textrm{max}] \}.
\label{eq:contact}
\end{equation}
When the MSCR is pushed within the lumen, the rod centerline of MSCR should be constrained by the tube,
 \begin{equation}
\forall s , \; \mathbf{r} (s) \in  \mathcal{S},
\end{equation}

\paragraph{Equilibrium equations} The equilibrium equations can be derived using functional variation,
\begin{equation}
\delta \mathcal{E}_{\mathrm{ela}} + \delta \mathcal{E}_{\mathrm{mag}} = \delta \mathcal{E}_{\mathrm{ext}},  
\end{equation}
where $\mathcal{E}_{\mathrm{ext}}$ is the work done by the external loadings.
The final version of the governing equations is the statement of force and moment balance under geometric constraints \citep{sano2022kirchhoff},
%\fix{can we connect this with previous equations?} \WH{it would be slightly hard, I think we can keep it}
\begin{equation}
\begin{aligned}
\mathbf{N}'(s) & + \mathbf{f}_{\mathrm{mag}} (s) + \mathbf{f}_{\mathrm{ext}} (s) = \mathbf{0} \\
\mathbf{M}' (s) + \mathbf{m}_{3} (s) \times \mathbf{N}' (s) & + \bm{\tau}_{\mathrm{mag}} (s) + \bm{\tau}_{\mathrm{ext}} (s) = \mathbf{0} \\
\textrm{s.t.} \; & \forall s , \; \mathbf{r} (s)  \in  \mathcal{S}.
\end{aligned}
\label{eq::EquilibriumEquations}
\end{equation}
where $\mathbf{f}_{\mathrm{ext}}$ (and $\bm{\tau}_{\mathrm{ext}}$) is the external force density (external moment density), $\mathbf{N}$ (and $\mathbf{M}$) is the internal elastic force (and internal elastic moment), and $\mathbf{f}_{\mathrm{mag}}$ (and $\bm{\tau}_{\mathrm{mag}}$) is the magnetic force density (magnetic moment density),
\begin{equation}
\begin{aligned}
\mathbf{N}(s) &= N_{1} (s) \mathbf{m}_{1} (s) + N_{2} (s) \mathbf{m}_{2} (s) + N_{3} (s) \mathbf{m}_{3} (s) \\
\mathbf{M}(s) &= M_{1}(s) \mathbf{m}_{1}(s) + M_{2}(s) \mathbf{m}_{2}(s) + M_{3}(s) \mathbf{m}_{3}(s) \\
\mathbf{f}_{\mathrm{mag}} (s) & = \bm{\mathcal{M}}(s) \cdot \nabla \mathbf{B}_{a} (s)\\
\bm{\tau}_{\mathrm{mag}} (s) & = \bm{\mathcal{M}} (s) \cross \mathbf{B}_{a} (s).
\end{aligned}
\label{eq::equilibrium}
\end{equation}
Note that the moment is linearly related to the curvature based on the linear constitutive law in Eq.~(\ref{eq:ConstitutiveEquations}), i.e., 
\begin{equation}
\begin{aligned}
M_{1}(s) &= EI \kappa_{1}(s) \\
M_{2}(s) &= EI \kappa_{2}(s) \\
M_{3}(s) &= GJ \kappa_{3}(s).
\end{aligned}
\label{eq::constitutiveLaw}
\end{equation}
By specifying the various boundary conditions, the configuration of the rod can be solved with the group of differential equations.
%
%\fix{Are these ordinary differential equations?} \WH{Yes.}
% The group of ordinary differential equations can be solved with a specified boundary condition.  

\section{Forward numerical simulation} \label{sec:simulation}

Directly solving Eqs.~(\ref{eq::EquilibriumEquations}),~(\ref{eq::equilibrium}), and~(\ref{eq::constitutiveLaw}) with constraints is hard and even impossible.
In this section, we present the design of our numerical framework, which is built upon the theoretical modeling presented in Section~\ref{sec:theory}, to simulate the dynamics of a MSCR navigating through confined lumens.

\begin{figure}[ht]
\centering
\includegraphics[width=\columnwidth]{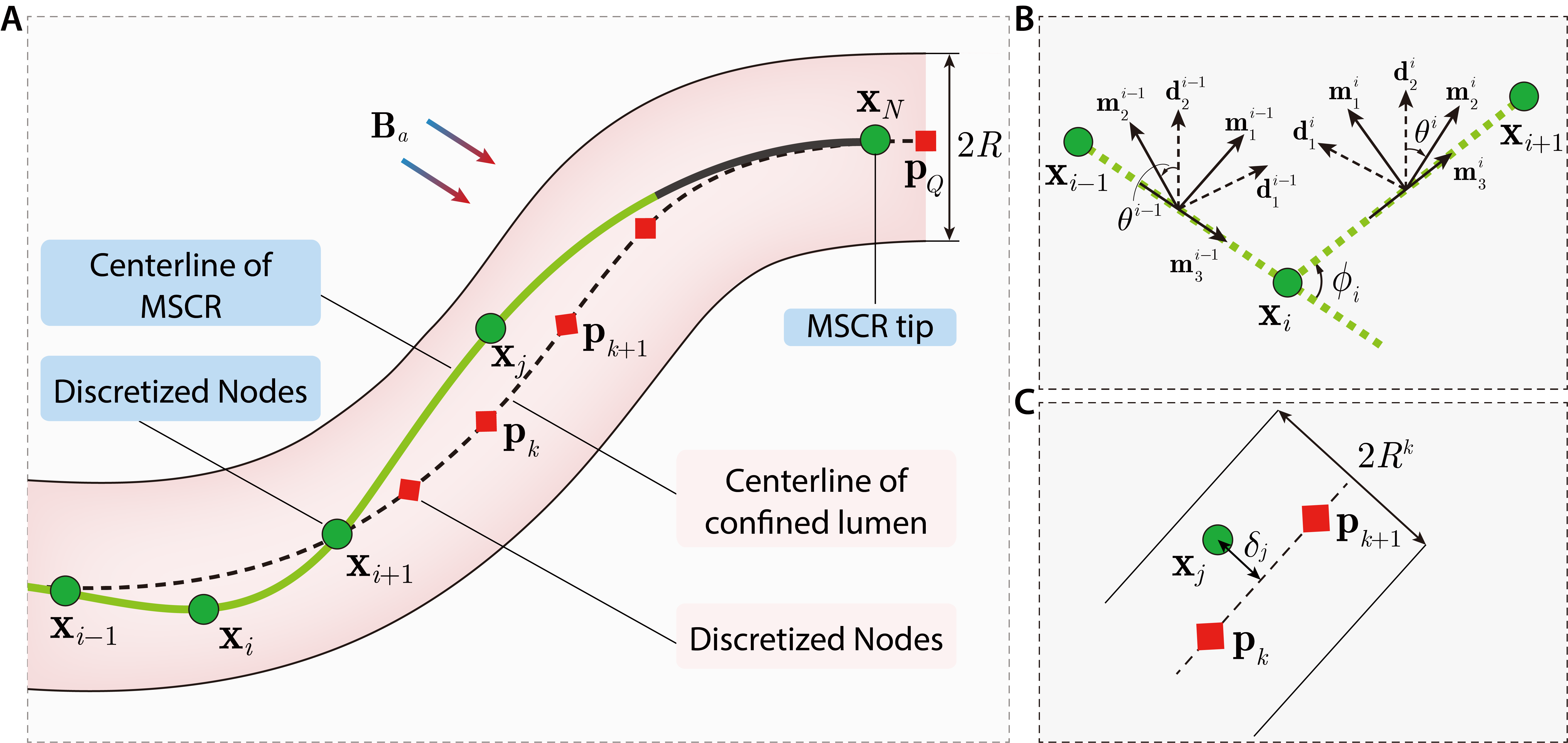}
\caption{Discretization of the MSCR and confined lumen. The MSCR is discretized as a magneto-elastic rod with the centerline path $s$, and the confined lumen is approximated as a cylindrical tube with the centerline path $S$. (A) Discrete diagram of rod (labeled as $\{ \mathbf{x}_1, \cdots, \mathbf{x}_{N} \}$) and tube (labeled as $\{ \mathbf{p}_1, \cdots, \mathbf{p}_{M} \}$). (B) Discrete measurements for bending and twisting of the rod. (C) Contact model between rod and tube, where $\delta_j$ measures the minimal distance from rod node $\mathbf{x}_j$ to the tube path $S$.}
\label{fig:model}
\end{figure}

\paragraph{Mechanics of MSCR} We use the discrete elastic rod (DER) algorithm to capture the geometrically nonlinear deformations of MSCR \citep{bergou2008discrete, bergou2010discrete}.
A slender rod-like MSCR can be described by its centerline and the associated material frame (Fig.~\ref{fig:model}(A)).
To capture its geometrically nonlinear configuration, the continuous body is discretized into $ N $ nodes,
\begin{equation}
\left[ \mathbf{x}_{1}, \mathbf{x}_{2}, ..., \mathbf{x}_{i}, ..., \mathbf{x}_{N} \right], \; \mathrm{with} \; i \in [1, N],
\end{equation}
which leads to $ N - 1 $ edge vectors, 
\begin{equation}
\left[ \mathbf{e}^{1}, \mathbf{e}^{2}, ..., \mathbf{e}^{N-1} \right],
\end{equation}
where 
\begin{equation}
\mathbf{e}^{i} = \mathbf{x}_{i+1} - \mathbf{x}_{i}, \; \mathrm{with} \; i \in [1, \ldots, N-1].
\end{equation}
Hereafter, subscripts are adopted for the quantities associated with nodes, e.g., $\mathbf x_i$, and superscripts are for the quantities associated with edges, e.g., $\mathbf e^i$.
Each edge, $ \mathbf{e}^{i} $, has an orthonormal adapted reference frame, $\left\{\mathbf{d}^{i}_{1}, \mathbf{d}^{i}_{2}, \mathbf{d}^{i}_{3} \right\}$, together with a material frame,
$ \left\{\mathbf{m}^{i}_{1}, \mathbf{m}^{i}_{2}, \mathbf{m}^{i}_{3} \right\}$,
and both of them share the tangential director, 
\begin{equation}
\mathbf{d}^{i}_{3} \equiv \mathbf{m}^{i}_{3}  = \frac {\mathbf e^i}  {|| \mathbf e^i ||}.
\end{equation}
The rotational difference between two frames along the tangential director is denoted by $\theta^{i}$, as shown in Fig.~\ref{fig:model}(B).
Nodal positions and twist angles constitute a $ 4N-1 $  degree-of-freedom (DOF) vector,  
\begin{equation}
\label{eq:dof}
\mathbf{q} = \left[\mathbf{x}_{1}, \theta^{1}, \mathbf{x}_{2}, ..., \mathbf{x}_{N-1}, \theta^{N-1}, \mathbf{x}_{N} \right],
\end{equation}
to describe the deformed configuration of a 1D rod-like object.
The strains of a deformed rod comprise three parts: stretching, bending, and twisting.
The discrete stretching strain of the $i$-th edge, $ \mathbf{e}^{i} $, can be evaluated as
\begin{equation}
\varepsilon^i = \frac{|| \mathbf e^i ||}{|| \bar{\mathbf{e}}^i ||} -1.
\label{eq:StrechingStain}
\end{equation}
Here, the quantity with a bar on top indicates its undeformed configuration, i.e., $|| \bar{\mathbf{e}}^i ||$ is the undeformed length of $i$-th edge vector. %\fix{I think we define this before in Section 2}
Similarly, we use 
\begin{equation}
\Delta \bar{l}_i =  \frac{1} {2} ( || \bar{\mathbf{e}}^i || + || \bar{\mathbf{e}}^{i-1}|| ) 
\end{equation}
as the Voronoi length associated with the $i$-th node.
Next, referring to Fig.~\ref{fig:model}(B), the bending curvature distribution can be evaluated by the curvature binormal that measures the misalignment between two consecutive edges at $\{ \mathbf{x}_{i-1}, \mathbf{x}_{i}, \mathbf{x}_{i+1} \}$ \citep{bergou2008discrete, bergou2010discrete},
\begin{equation}
\begin{aligned}
\kappa_{1,i} & =  \frac { {\left( \mathbf m_2^{i-1} + \mathbf m_2^i \right) } } {\Delta \bar{l}_i } \cdot \frac {\left( \mathbf{e}^{i-1} \times \mathbf{e}^{i} \right) }  { || \mathbf{e}^{i-1} || \cdot || \mathbf{e}^{i} || + ||\mathbf{e}^{i-1} \cdot \mathbf{e}^{i}|| } \\
\kappa_{2,i} & = - \frac { {\left( \mathbf m_1^{i-1} + \mathbf m_1^i \right) } } {\Delta \bar{l}_i } \cdot \frac {\left( \mathbf{e}^{i-1} \times \mathbf{e}^{i} \right) }  { || \mathbf{e}^{i-1} || \cdot || \mathbf{e}^{i} || + ||\mathbf{e}^{i-1} \cdot \mathbf{e}^{i}|| },
\end{aligned}
\label{eq:bendingStrain}
\end{equation}
The twisting strain distribution at the $i$-th node is measured as,
\begin{equation}
\kappa_{3,i} = \frac  {({ \theta^{i} - \theta^{i-1} + {m}_{i}^{\mathrm{ref}} } )} {\Delta \bar{l}_i } ,
\label{eq:TwistingCurvature}
\end{equation}
where $ {m}_{i}^{\mathrm{ref}} $ is the reference twist associated with the reference frame and can be computed from a parallel transport via time marching \citep{jawed2018primer}.
%s
The total elastic potentials of a rod system are the sum of the stretching, bending, and twisting, expressed in its discretized form in terms of the DOF vector $\mathbf{q}$, as
\begin{equation}
E^{\mathrm{ela}} = ( E^{\mathrm{s}} + E^{\mathrm{b}} + E^{\mathrm{t}}),
\end{equation}
where
\begin{equation}
\begin{aligned}
E^{\mathrm{s}} &=  \sum_{i=1}^{N-1} \frac{1}{2}EA \left( \varepsilon^i \right)^2 || \bar{\mathbf{e}}^i ||, \\
E^{\mathrm{b}} &= \sum_{i=1}^{N} \frac{1}{2}  {EI}  (\kappa_{1,i})^2  \Delta \bar{l}_i + \sum_{i=1}^{N} \frac{1}{2}  {EI} (\kappa_{2,i})^2 \Delta \bar{l}_i , \\
E^{\mathrm{t}} &= \sum_{i=1}^{N} \frac{1}{2} {GJ} (\kappa_{3,i} )^2 {{\Delta \bar{l}_i} } .
\label{eq:discreteEnergies}
\end{aligned}
\end{equation}
The elastic force can then be computed as the gradient of the elastic energy,
\begin{equation}
\mathbf{F}^{\mathrm{e}} = - \frac {\partial E^{\mathrm{ela}} } {\partial \mathbf{q}}.
\end{equation}

\paragraph{Magnetic actuation} The MSCR with magnetization $\bm{\mathcal{M}}$ can be actively actuated by the actuation magnetic field $\mathbf{B}_{a}$.
The discrete format of the magnetic functional is \citep{huang2023modeling, huang2023discrete}
\begin{equation}
E^{\mathrm{mag}} = \sum_{i=1}^{N-1} -\left( \bm{\mathcal{M}}^{i} \cdot \mathbf{B}_{a} \right) || \bar{\mathbf{e}}^i ||,
\label{eq:discreteMagneticEnergies}
\end{equation}
where $\bm{\mathcal{M}}^{i}$ is the magnetization vector of $i$-th edge,
\begin{equation}
\bm{\mathcal{M}}^{i} = \left[ \left( \mathbf{m}_{1}^{i} \otimes \bar{\mathbf{m}}_{1}^{i} + \mathbf{m}_{2}^{i} \otimes \bar{\mathbf{m}}_{2}^{i} + \mathbf{m}_{3}^{i} \otimes \bar{\mathbf{m}}_{3}^{i} \right) \cdot \bar{\bm{\mathcal{M}}^{i}} \right].
\end{equation}
Finally, the external magnetic force vector can be evaluated in a manner similar to the elastic force vector,
\begin{equation}
\mathbf{F}^{\mathrm{m}} = - \frac {\partial E^{\mathrm{mag}} } {\partial \mathbf{q}}.
\end{equation}

\paragraph{Contact with a lumen} The MSCR is constrained by a lumen such that the contact constraint needs to be included in the numerical model. The confined lumen is approximated as a curved tube
in which the centerline can be described as $M$ nodes,
\begin{equation}
\left[ \mathbf{p}_{1}, \mathbf{p}_{2}, ..., \mathbf{p}_{k}, ..., \mathbf{p}_{Q} \right], \; \mathrm{with} \; k \in [1, Q],
\end{equation}
and the local tube radius between the $k$-th and $(k+1)$-th node is defined as $R^{k}$, which is used for contact defection and constrained simulation. %\fix{Do we really use $R_k$ anywhere?}
Note that the tube centerline is assumed to be fixed during the simulation process.
The contact element is constructed based on the minimum distance between each node and the tube centerline, and $3$ nodes are considered here, i.e., $ \{ \mathbf{x}_{j}, \mathbf{p}_{k}, \mathbf{p}_{k+1} \} $, where $\mathbf{p}_{k}$ and $\mathbf{p}_{k+1}$ are the vertices on the tube centerline, and $\mathbf{x}_{j}$ represents $j$-the node on the MSCR, referring to Fig.~\ref{fig:model}C.
The minimum distance between the $j$-th node and the tube centerline is defined as $ \delta_{j} $,
\begin{equation}
\delta_{j} = \mathrm{min} \left \{ ||\mathbf{x}_{j} - \mathbf{p}_{k}||, ||\mathbf{x}_{j} - \mathbf{p}_{k+1}||, \frac {||(\mathbf{x}_{j} - \mathbf{p}_{k}) \times (\mathbf{x}_{j} - \mathbf{p}_{k+1})||} {||\mathbf{p}_{k} - \mathbf{p}_{k+1}||}  \right \}.
\end{equation}
To ensure the non-deviation condition between the node and the tube centerline, the following barrier energy function is constructed by \citep{choi2021implicit, tong2023fully, huang2024dynamic}
\begin{equation}
C_{j} = 
\begin{cases}
- K_{c} \left[ (d_{j} - \tilde{d})^2 \log (  d_{j} / \tilde{d} )  \right] & \; \mathrm{ when } \; 0 < d_{j} < \tilde{d} \\
0 & \; \mathrm{ when } \; d_{j} \ge \tilde{d}
\end{cases},
\label{eq::contactPotential}
\end{equation}
where $K_{c}$ is the stiffness parameter, $\tilde{d}$ is a barrier parameter, and $d_{j} = R^{k} - \delta_{j} $, thus the total contact potential in the sum of contact set $\mathcal{C}$ constructed based on continuous contact detect,
\begin{equation}
E^{\mathrm{con}} = \sum_{j \in \mathcal{C}} C_{j}.
\end{equation}
The contact force is the minus gradient of the contact potential, which is expressed as
\begin{equation}
\mathbf{F}^{\mathrm{c}} = - \frac {\partial E^{\mathrm{con}} } {\partial \mathbf{q}}.
\end{equation}

\paragraph{Equations of motion} Overall, we consider the inertial and damping effects and use a dynamic approach to solve the nonlinear equations.
At the $k$-th time step, $t_{k}$, the DOF vector $\mathbf{q}$ can be updated from $t_{k}$ to $t_{k+1} = t_k + dt$ by imposing the $(4N-1)$-sized equation of motion,
%\fix{is this displacement?} \WH{should be DOF vector}
\begin{equation}
\begin{aligned} 
\mathcal{DER}  \equiv    \mathbb{M}  \ddot{\mathbf{q}}(t_{k+1}) &- \mu \mathbb{M} \dot{\mathbf{q}}(t_{k+1}) - d t \; \left[ \mathbf{F}^{\text{e}}(t_{k+1}) + \mathbf{F}^{\text{m}}(t_{k+1}) + \mathbf{F}^{\text{c}}(t_{k+1}) \right] = 0 \\
 \mathbf{q}(t_{k+1})  &=   {\mathbf{q}}(t_{k}) + d t \; \dot{\mathbf{q}}(t_{k+1})  \\
\dot{\mathbf{q}}(t_{k+1})&=  \dot{\mathbf{q}}(t_{k}) + d t \; \ddot{\mathbf{q}}(t_{k+1}) ,
\end{aligned}
\label{eq:implicitEuler}
\end{equation}
where $\mathbb{M}$ is the diagonal mass matrix comprised of lumped masses, $\mu$ is the damping coefficient, and $ d t $ is the time step size.
%
% Note that the DOF vector and the reaction force satisfy a complementary condition, i.e., the reaction force is zero when the associated DOF is free and the reaction force is nonzero when the associated DOF is prescribed. 
%
The iterative Newton-Raphson method is employed to solve the nonlinear equations of motion derived in Eq.~(\ref{eq:implicitEuler}).

\section{Model-based Control Framework} \label{sec:control}

% The use of MSCR in intervention procedures presents two major challenges related to contact between its distal tip and the lumen walls: (1) increased risk of puncturing fragile walls, potentially endangering the patient and complicating the procedure, and (2) impeded smooth navigation, as contact with the walls can hinder movement in confined environments.
%
In this section, we define the control problem aimed at ensuring the safe and real-time guidance of a MSCR within confined lumens.
The objective is to derive the optimal external magnetic field so that the MSCR's tip can align with the lumen's centerline when the other end of MSCR is pushed with the velocity $v_0$ into the lumen.
In other words, the terminal node of the MSCR should follow the lumen centerline, which results in the following constrained optimization problem:
\begin{equation} 
\begin{aligned}
{\mathbf{B}}_{a}^* &= \arg \min_{{\mathbf{B}}_{a}} \ \lVert \delta_N \rVert \\
\textrm{s.t.} \quad 
& \mathcal{DER}(\mathbf{q}, \mathbf B_a) = \mathbf{0}, \\
%& \dot{\mathbf q_e} = v_0 \mathbf P(0)',
\end{aligned} 
\label{eq::optimization} 
\end{equation}
where ${\mathbf{B}}_{a}^* \in \mathcal{R}^{3\times1} $ is the optimal actuation magnetic field, $\mathcal{DER}(\mathbf q, \mathbf B_{a})$ describes the mechanical response of the MSCR in Eq.~(\ref{eq:implicitEuler}), 
%and $\mathbf q_e$ is the discrete node which is about to enter into the lumen, which defines the boundary conditions for $\mathcal{DER}$.
In principle, Lagrange multipliers can be introduced for the nonlinear equality constraint to form a Karush–Kuhn–Tucker (KKT) system whose stationarity and feasibility must be satisfied~\citep{ruszczynski2011nonlinear}.
However, the high dimensionality of the equations of motion $\mathcal{DER}(\mathbf q, \mathbf B_{a})$ and the embedded inequality contact constraints in $\mathcal{DER}(\mathbf q, \mathbf B_{a})$ require active-set updates, which are numerically sensitive and expensive. %\fix{we can delete this} 

\begin{figure}[!h]
\centering
\includegraphics[width=0.9\columnwidth]{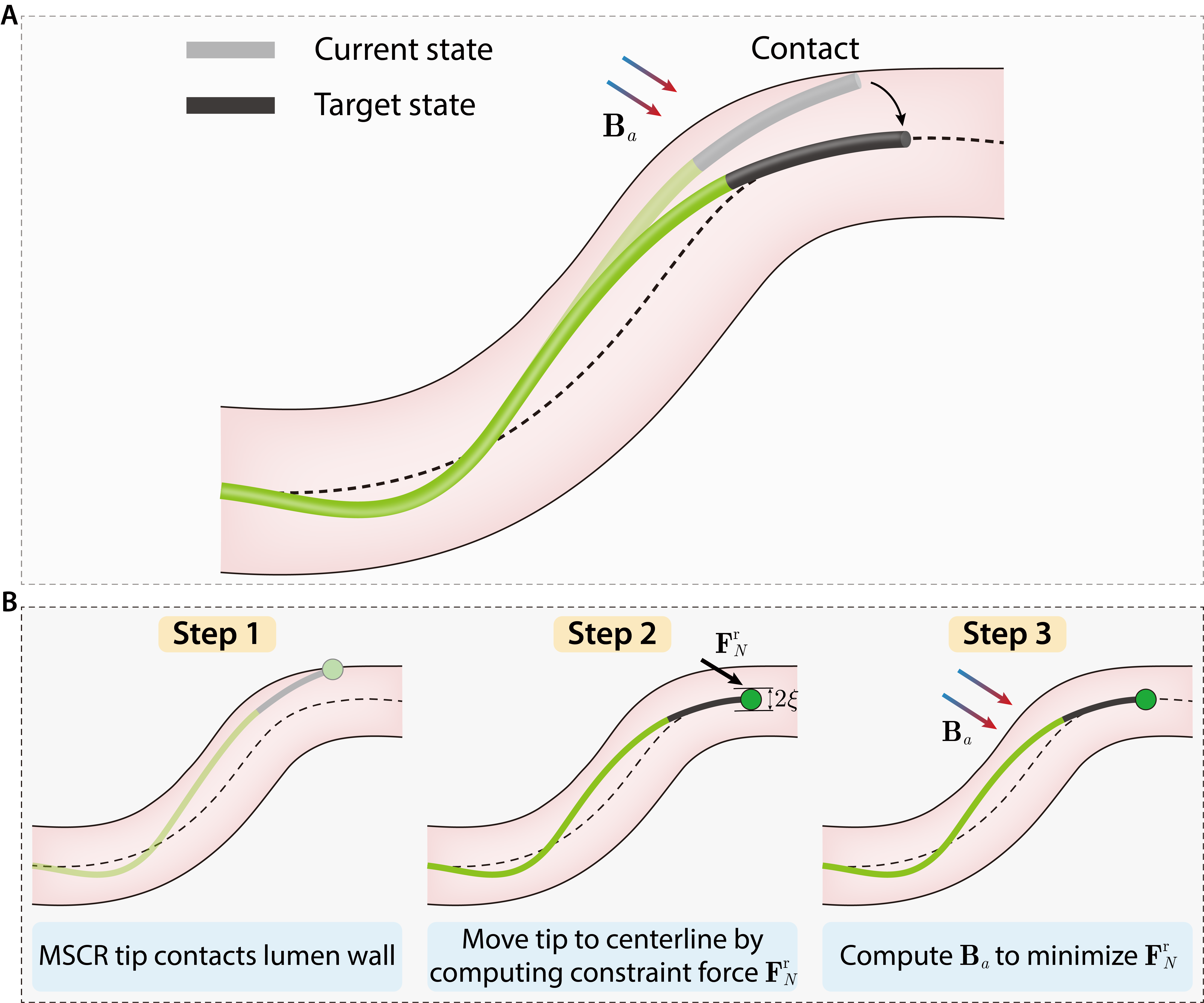}
\caption{Demonstration of model-based control. (A) Magnetic field control prevents the rod tip from contacting the lumen, highlighting precise manipulation. (B) Control process: Step 1 — detect contact between the rod tip and the lumen; Step 2 — virtually relocate the rod tip to a contact-free region and compute the equivalent constraint force; Step 3 — determine the optimal magnetic field strength to minimize the constraint force and apply it to the current configuration.}
\label{fig:control}
\end{figure}

To circumvent these difficulties, we propose a model-based control method that leverages the proposed reduced-order simulations and embeds the optimization objective in Eq.~(\ref{eq::optimization}) into the equilibrium equation itself.
Although perfect alignment between the MSCR's tip and the lumen's centerline implies $\delta_N = 0$. Here, we slack this equality constraint to an inequality:
\begin{equation}
\delta_{N} \leq \xi,
\label{eq:inequality}
\end{equation}
As shown in Fig.~\ref{fig:control}(B), Eq.~(\ref{eq:inequality}) describes a small ``virtual tube'' of radius $\xi \ll R$ around the lumen centerline, where $R$ is the local radius of the lumen.
To enforce $\delta_N \leq \xi$ without resorting to a direct inequality in the KKT sense, we utilize the contact potential stated in Eq.~(\ref{eq::contactPotential}) to compute constraint force applied on the last node $\mathbf{F}_{N}^{\text{r}} \in \mathcal{R}^{3 \times 1}$.
If the tip remains within $\xi$, the constraint force $\mathbf F_N^{\text{r}} \approx \mathbf{0}$. However, as the tip approaches the boundary of the virtual tube, the constraint force, $\mathbf F_N^{\text{r}}$, would ``push'' it back, effectively encoding the inequality constraint Eq.~(\ref{eq:inequality}) in the mechanics.
Our control objective is to determine the optimal actuation magnetic field $\mathbf B_a^{*} $ that keeps the tip aligned with the tube centerline without exerting any significant constraint force on the small virtual tube.
Note that,
In other words, this model-based analysis can lead to the following objective function:
\begin{equation} 
\begin{aligned}
{\mathbf{B}}_{a}^* &= \arg \min_{{\mathbf{B}}_{a}} \ \lVert \mathbf F_N^r \rVert \\
\end{aligned}
\label{eq::control} 
\end{equation}
with the following equality constraint:
\begin{equation}
 {\mathcal{DER}}(\mathbf{q}, \mathbf B_a) + \mathbf F^r(\mathbf B_a) = \mathbf{0}, 
%& \dot{\mathbf q_e} = v_0 \mathbf P(0)'.
\label{eq:updatedDER}
\end{equation}
where $\mathbf F^r \in \mathcal R^{(4N-1) \times 1}$ is the constraint force and only non-zero elements are $\mathbf F_N^r$ applied on the last node.
Note that the constraint force $\mathbf F_N^r $ can be computed by solving the updated EOMs stated in Eq.~(\ref{eq:updatedDER}).
Therein, we can view Eq.~(\ref{eq::control}) as a root-finding problem on constraint force $\mathbf F_N^{\text{r}}$:
\begin{equation} 
\begin{aligned} 
\mathbf{F}_{N}^{\text{r}} (\mathbf{B}_{a}) = \mathbf{0}.
\end{aligned} 
\label{eq::root} 
\end{equation}
To effectively explore the root of Eq.~(\ref{eq::root}), we employ a finite difference method to compute the numerical Jacobian:
\begin{equation}
\mathbb {J}^\textrm{opt} = \left[ 
\frac{\mathbf F_{N}^{\text{r}}( {B}_{x} + \delta e) - \mathbf F_{N}^{\text{r}}( {B}_{x}) }{\delta e},
\frac{\mathbf F_{N}^{\text{r}}( {B}_{y} + \delta e) - \mathbf F_{N}^{\text{r}}( {B}_{y}) }{\delta e},
\frac{\mathbf F_{N}^{\text{r}}( {B}_{z} + \delta e) - \mathbf F_{N}^{\text{r}}( {B}_{z}) }{\delta e} \right]
\label{eq::jacobian}
\end{equation}
where $ [{B}_{x}, {B}_{y}, {B}_{z}] \equiv \mathbf{B}_{a} $ is the components of the actuation magnetic field, $\delta e$ is the small perturbations number. With the computed $\mathbb {J}^\textrm{opt} \in \mathcal{R}^{3 \times 3}$, the Newton search step for updating the actuation magnetic field $\mathbf{B}_{a}$ can be computed with the momentum-based optimizer:
\begin{equation}
\mathbf{B}_{a}^{(n+1)} = \mathbf{B}_{a}^{(n)} - \beta_{1} \Delta \mathbf{B}_{a}^{(n)} - \beta_{2} \Delta \mathbf{B}_{a}^{(n+1)},
\end{equation}
where 
\begin{equation}
\Delta \mathbf{B}_{a}^{(n+1)} = (\mathbb {J}^\textrm{opt}) \backslash \mathbf F_{N}^{\text{r}}.
\end{equation}
Here, $\beta_1$ and $\beta_2$ are momentum coefficients.
When the constraint force $\mathbf F_{N}^{\text{r}}$ is smaller than a threshold, we derive the optimal corresponding external magnetic field $\mathbf B_a^*$ and the corresponding updated DOF vector $\mathbf{q}$, then we move to next time step and continue the pushing process of the MSCR. Thus, the active control process is built within the forward simulation framework.
The full algorithm for the model-based control framework is stated in Algorithm~\ref{algo:control}.

\begin{algorithm}
\caption{Model-based Control Framework} \label{algo:control}
\begin{algorithmic}[1]

\Require{$\mathbf q(t_k), \mathbf B_a^*(t_{k})$} \Comment{Current DOFs, actuation magnetic field}

\Require{$\xi, \beta_1, \beta_2$}  \Comment{$\xi$: virtual tube's radius; $\beta_1, \beta_2$: momentum coeffcients}

\Ensure{$\mathbf B_a^*(t_{k+1})$, $\mathbf q(t_{k+1})$}  \Comment{Optimal magnetic field and updated DOFs of MSCR}

\State{$n \gets 0$}

\State {$\mathbf B_a^{(n)} \gets \mathbf B_a^*(t_{k})$}
\Comment{Initialize the guess for actuation magnetic field}

\State {$ \left \{\mathbf F_N^{\text{r},(n)}, \mathbf q^{(n)} \right\} \gets \mathcal{DER}\left( \mathbf q(t_k), \mathbf B_a^{(n)} \right) + \mathbf F^{\text{r}, (n)}$} 
\Comment{Solve Eq.~(\ref{eq:updatedDER})}

\State {$\Delta \mathbf B_a^{(n)} \gets \mathbf 0$}

\While{$ \lVert \mathbf F_N^{\text{r},(n)} \rVert >$ tolerance}
\State{$\mathbb{J}^\textrm{opt} \gets$ Eq.~(\ref{eq::jacobian})}

\State{$\Delta \mathbf B_a^{(n+1)} \gets \mathbb{J}^\textrm{opt} \backslash \mathbf F_N^{\text{r},(n)}$}

\State{$\mathbf{B}_{a}^{(n+1)} \gets \mathbf{B}_{a}^{(n)} - \beta_{1} \Delta \mathbf{B}_{a}^{(n)} - \beta_{2} \Delta \mathbf{B}_{a}^{(n+1)}$}

% \State{$\Delta \mathbf B_a^{(n)} \gets \Delta \mathbf B_a^{(n+1)}$} \fix{reverse?}

\State{$n \gets n + 1 $}

\State {$ \left\{\mathbf F_N^{\text{r},(n)}, \mathbf q^{(n)} \right\} \gets \mathcal{DER}\left( \mathbf q(t_k), \mathbf B_a^{(n)} \right) + \mathbf F^{\text{r},(n)}$} 
\EndWhile

\State{$\mathbf{q}(t_{k+1}) \gets \mathbf{q}^{(n)}$}

\State{$\mathbf B_a^*(t_{k+1}) \gets \mathbf B_a^{(n)}$}

\State \Return {$\mathbf q(t_{k+1}), \mathbf B_a^*(t_{k+1})$}
\end{algorithmic}
\end{algorithm}

\section{Results} \label{sec:results}

In this section, we present numerical and experimental results to validate our modeling and control framework across different navigation scenarios. We begin by detailing the numerical setup and non-dimensional analysis for parameter normalization. Using the same normalized parameters, we then describe the experimental setup, ensuring consistency between simulations and physical validation. 
Next, we analyze passive navigation, where the MSCR is solely pushed by the advancer without an actuation magnetic field. The results reveal increased contact forces and trajectory deviations, demonstrating the limitations of passive control and emphasizing the need for an active control strategy.
We then evaluate our model-based control method, showing how it dynamically adjusts the actuation magnetic field to guide the MSCR through confined lumens of varying geometries. The results highlight the framework’s ability to minimize contact forces, maintain stable tip alignment, and ensure precise path tracking, even in complex lumen structures. %
Finally, we assess the computational efficiency of our approach, demonstrating that the reduced-order numerical model enables real-time updates of the actuation magnetic field, ensuring feasibility for medical and industrial applications. The findings confirm that our method provides an efficient and robust solution for safe and precise MSCR navigation in confined environments.

\begin{figure}[!h]
\centering
\includegraphics[width=0.8\columnwidth]{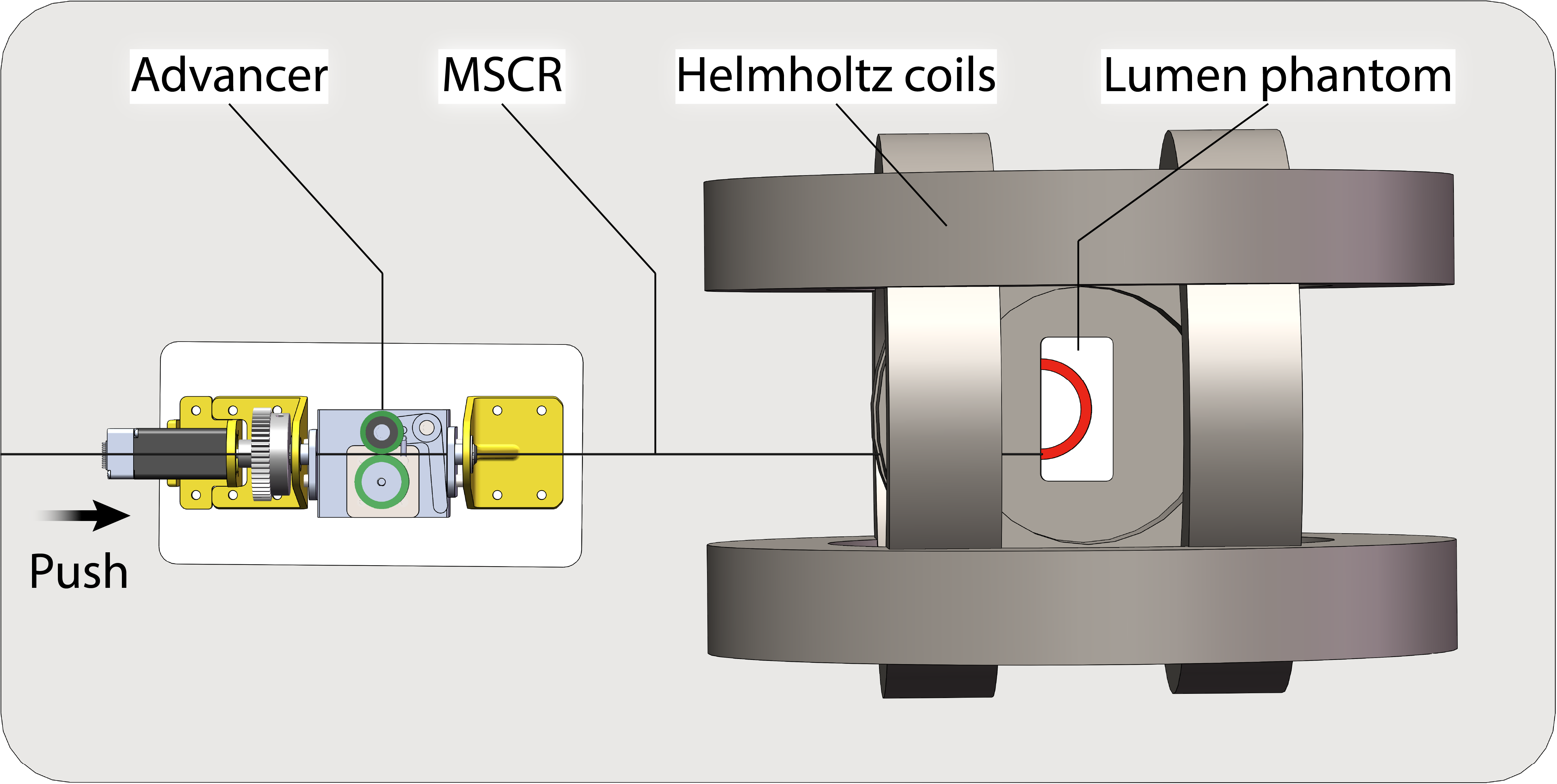}
\caption{Description of the experiment platform. The lumen phantom is placed in the center of the Helmholtz coils. An advancer clamps the MSCR and then pushes it to navigate the phantom path under the actuating magnetic fields.}
\label{fig:Exp}
\end{figure}

\subsection{Setup and normalization}
\paragraph{Experimental setup}
% \fix{Make simulation and experiment parameters consistent}

Here we choose a MSCR with a magnetized tip length of $L_{m} = 5.0$ mm, rod radius $r_{0} = 0.6$ mm, and total length $L = 100$ mm.
The magnetized tip of MSCR was fabricated by dispersing NdFeB particles (average diameter $\sim 5$ um) in the PDMS matrix (base-to-curing agent mass ratio $10:1$, Sylgard $184$, Dow Corning, USA). The Young's modulus and shear modulus of the magnetic composite was measured as $E = 3.0$ MPa and $G=1.0$ MPa, respectively.
The MSCR tip was magnetized along its axial direction using impulsed magnetic fields ($\sim 4$ T) generated by a digital pulse magnetizer, resulting in a remanent magnetization $ || \bar{\bm{\mathcal{M}}} || = 0.1448$ A$\cdot$m along the axial direction.
As shown in Fig.~\ref{fig:Exp}, the experiments were conducted in 3D-printed lumen phantoms placed at the center of Helmholtz coils. The MSCR was pushed by a stepper motor-driven advancer at a controlled speed. The magnitude and direction of actuation magnetic fields were controlled by adjusting the coil currents along three axes. To ensure consistency with the non-dimensional analysis, all applied magnetic fields were normalized for comparative evaluation.

\paragraph{Non-dimensional analysis}

To enhance the generalizability of our results and enable broader applicability to various analytical scenarios and provide a side-by-side comparison between the experimental results and numerical predictions, we employ Buckingham's PI theorem to normalize key variables, %\fix{I think there should be a $\int$ for $\mathcal{X}$?}
\begin{equation}
\begin{aligned} 
\bm{\mathcal{B}} & = { \mathbf{B}_{a} || \bar{\bm{\mathcal{M}}} ||   }L^2/{EI}, \\
\mathcal{X} & = {v_0  t} / {L}, \\
\mathcal{F} &=  || \mathbf F_{N}^{\text{r}} ||  {L^2} / {EI}.
\end{aligned}
\label{eq::mapping}
\end{equation}
where $\bm{\mathcal{B}} \equiv [\mathcal{B}_{x}, \mathcal{B}_{y}, \mathcal{B}_{z}]$ is the normalized magnetic field, $\mathcal{X}$ is the normalized push distance, and $\mathcal{F}$ is the normalized tip contact force.

\begin{figure}[!b]
\centering
\includegraphics[width=1\columnwidth]{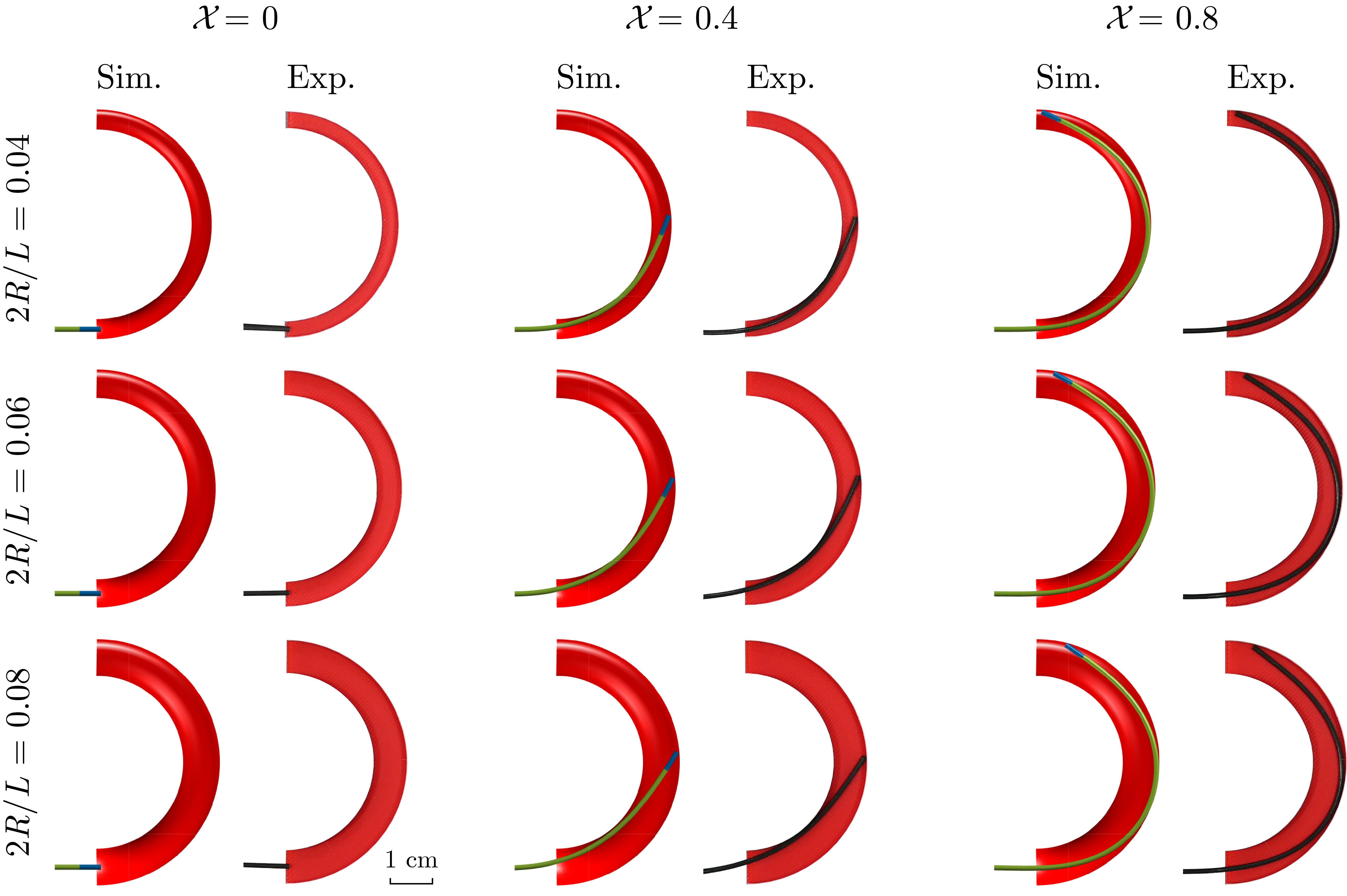}
\caption{Configurations of MSCR being pushed in half-circle lumen phantoms without magnetic fields in the simulation and experiment, respectively. From top to bottom, the lumen phantom diameters are varied as $2R/L \in \{0.04, 0.06, 0.08 \}$.}
\label{fig:planar_validation}
\end{figure}

\paragraph{Numerical setup}
In our numerical analysis, initially, all discretized nodes and edges are fixed and manually moved at a constant speed, $v_0$, along the $x$-axis, which is defined as the tangential direction at the lumen entrance, until they enter the lumen.
Once inside, the constraints are released, allowing the nodes and edges to evolve freely according to the dynamic equilibrium conditions.
The physical and geometric parameters of the MSCR in the simulation are selected to match the experimental conditions to ensure a fair evaluation of the model-based control framework in real-world experiments.
%
% The following physical and geometric parameters are used: rod length $L=1.0$ m, magnetic tip length $L_{m} = 5.0$ cm, rod radius $r_{0} = 0.1$ mm, Young's modulus $E=1.0$ GPa, shear modulus $G=0.33$ GPa, material density $\rho = 1.0 \; \si{g/cm^3} $, damping coefficient, $\mu=0.01$, and rod tip magnetization density ${\mathcal{M}} = 100$ kA/m, aligned with the rod tangential direction.
%
% Even though the numerical inputs are different from the exact experimental data, we keep all non-dimensional parameters identical after normalization.
%
Also, in numerical study, we adopt the following parameters based on a convergence study: total rod nodal number $N=100$, total lumen phantom nodal number $Q=200$, time step size $dt=10$ ms, contact stiffness $K_{c} = 1.0$ kPa, contact barrier $\tilde{d}= 0.1$ mm, virtual tube radius $\xi = 0.01 R$, and insert speed $v_{0} = 1.0 \; \si{mm/s}$.
The insertion speed is relatively low compared with the elastic wave to ensure the quasi-static analysis.
For model-based control, we use a perturbation
magnitude $\delta e = 10^{-4}$ to compute the numerical Jacobian, as well as optimization rate $\beta_{1} = 0.9$ and $\beta_{2} = 10^{-4}$ for momentum-based gradient optimizer.

% \subsection{Variable Normalization}

% \label{subsec:val}
% \begin{figure}[ht]
% \centering
% \includegraphics[width=1\columnwidth]{Figures/fig_planar.png}
% \caption{Simulation of rod-tube interaction. (A) MSCR passes through half-circle tunnels with the same radius of 0.2\si{m} but varying thicknesses: Case 1 -- 0.01\si{m}; Case 2 -- 0.02\si{m}; Case 3 -- 0.03\si{m}; Case 4 -- 0.04\si{m}. (B) Normalized distal contact force plotted against normalized insertion distance for various cases. \fix{ Wangliu: only three cases. Spiliting A and B, change case 1 to make, B. the outlines should be thinner. adding more tickles on y axis}}
% \label{fig:planar}
% \end{figure}

% To enhance the generalizability of our results and enable broader applicability to various analytical scenarios, we introduce a normalization framework for key analytical variables, including force, distance, and magnetic field strength. Normalization ensures that these variables are dimensionless and scaled consistently, allowing comparisons between different systems and conditions. 

% The distal contact force is normalized as
% \begin{equation}
% \mathcal{F} =  \frac {RL^2} {EI},
% \end{equation}
% leawhere $R$ is the magnitude of the $\mathbf{R}$.
% %
% The normalized insertion distance is defined as 
% \begin{equation}
% \mathcal{X} = \frac {v_{0} t} {L}.
% \end{equation}

% The actuation magnetic field is normalized as
% \begin{equation}
% \mathcal{B} = \frac {B_{a} \mathcal{M}  L^2} {EI}.
% \end{equation}
% where $\mathcal{M}$ (and $B_{a}$) is the magnitude of the  $\bm{\mathcal{M}}$ (and $\mathbf{B}_{a}$).

\subsection{Passive pushing of the MSCR}

In this subsection, we examine a case where the MSCR navigates under a passive pushing without magnetic actuation.
Two planar cases are considered here: (i) a half-circle lumen phantom, and (ii) a sinusoidal lumen phantom, as both of them are commonly appear in human arteries and veins. 
We first consider a MSCR moving in a half-circle lumen phantom.
The lumen phantom diameter varied in $2R/L \in \{0.04, 0.06, 0.08 \}$, and centerline radius is selected as $\hat{R}_{c} / L = 0.25$, i.e., the centerline is given by %\fix{where is $R_c$'s definition}
\begin{equation}
\begin{cases}
 {\hat{X}_{c}} / L &= 0.25 \cos(\omega), \\
 {\hat{Y}_{c}} / L &= 0.25 \sin(\omega) + 0.25,
 \end{cases}
\;\;\mathrm{with} \;  \omega \in [0, \pi].
\end{equation}
The comparison between the experiments and the simulation is presented in Fig.~\ref{fig:planar_validation}.
The excellent agreement between them validates the physical accuracy of our proposed contact simulation.

\begin{figure}[!h]
\centering
\includegraphics[width=1\columnwidth]{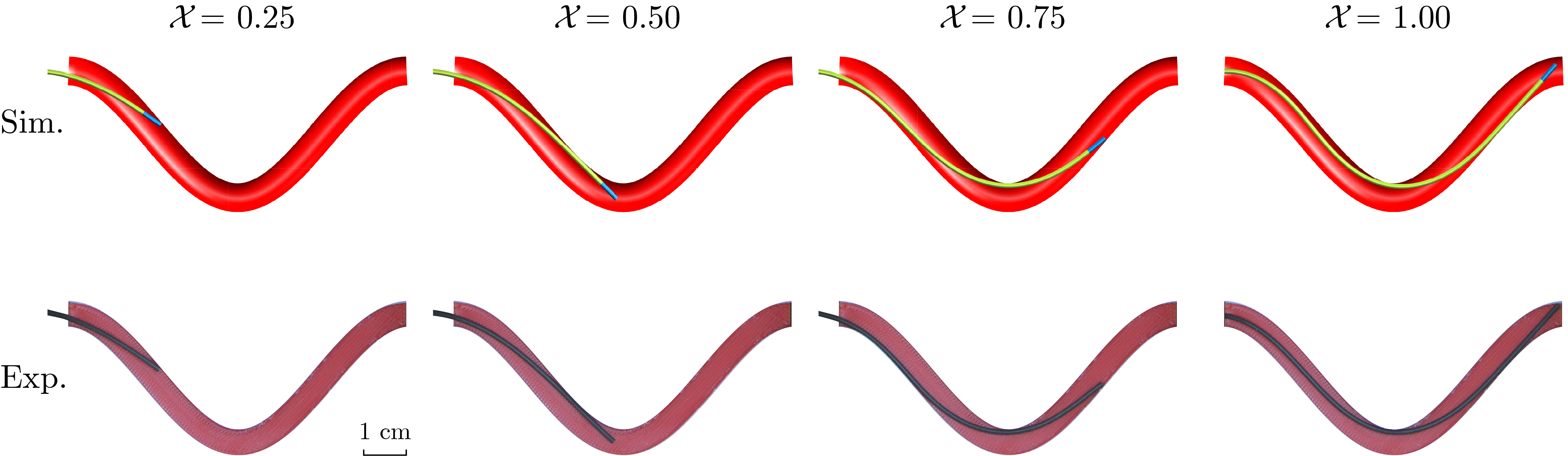}
\caption{Configurations of MSCR being pushed in a sinusoidal lumen phantom without magnetic fields in the simulation and experiment, respectively. Here, the lumen phantom diameter is $2R/L =0.06$.}
\label{fig:planar_validation_2}
\end{figure}

Next, a similar scenario for the MSCR moving in a sinusoidal lumen phantom is provided in Fig.~\ref{fig:planar_validation_2}.
Here, the lumen phantom diameter is fixed as $2R/L =0.06$ and its centerline can be described as,
\begin{equation}
 {\hat{Y}_{s}} /L  = 0.15 \cos( {\pi} { \hat{X}_{s}} / {40 L}) - 0.15, \; \mathrm{with} \; \hat{X}_{s}/L \in [0, 80].
\end{equation}
Physical contact between the MSCR and the lumen phantom can be observed and the configuration of the MSCR is determined by both the push distance and the lumen phantom shape.
Again, qualitative agreements between the numerical prediction and the experimental observation indicate the accuracy of our discrete model.

\subsection{Active Control of the MSCR Navigating in Planar lumen phantoms}

In this subsection, we apply the proposed model-based control algorithm to compute the actuation magnetic field required to navigate the MSCR through the two-dimensional (2D) planar lumen phantoms discussed before: a half-circle and a sinusoidal lumen phantom.
We conduct a comprehensive study on how the distal tip contact force varies during navigation to illustrate the effectiveness of the model-based control algorithm. 
Additionally, desktop experiments are performed to further validate the proposed numerical framework and active control method.

\begin{figure}[!h]
\centering
\includegraphics[width=1\columnwidth]{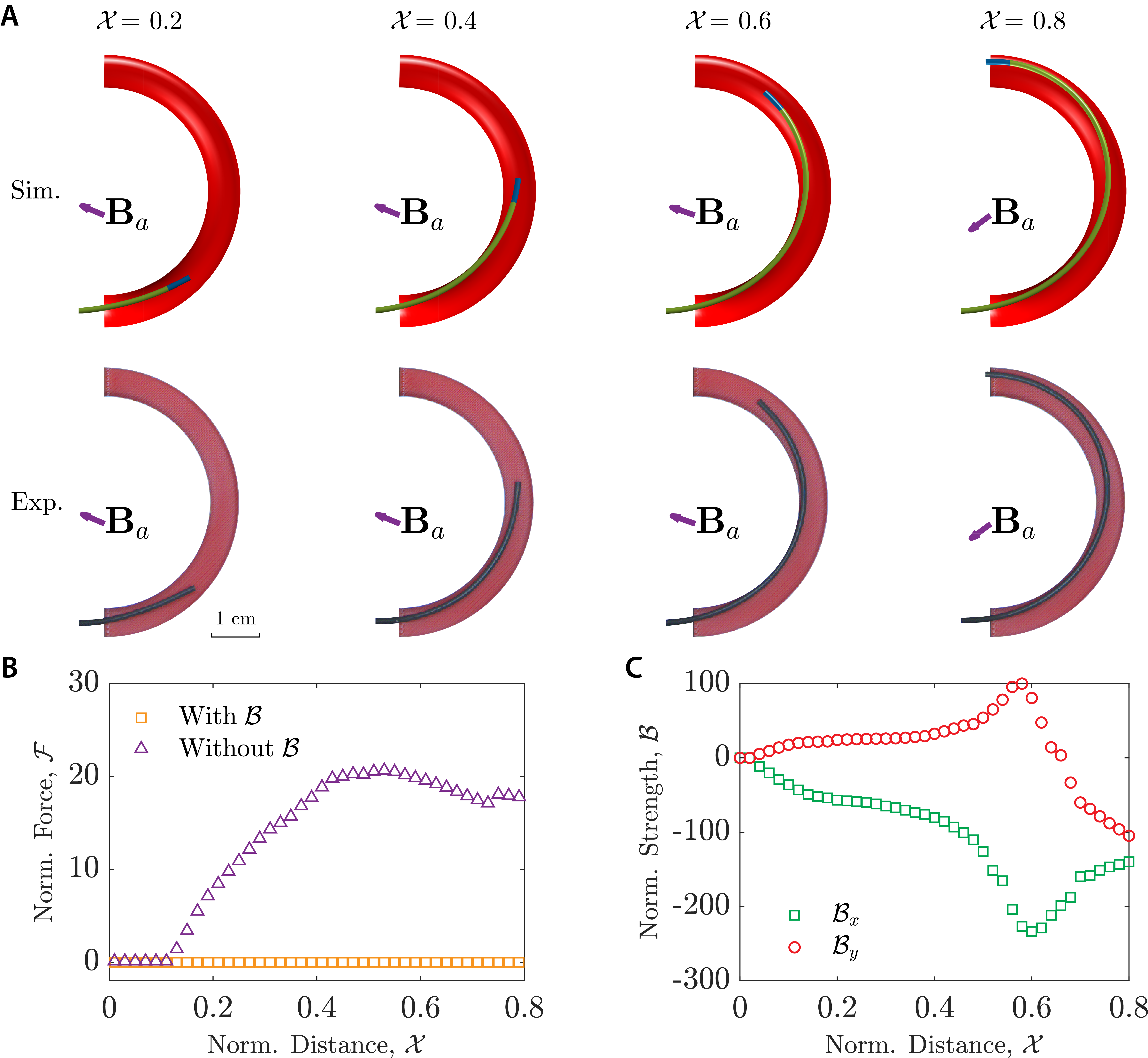}
\caption{Navigation of MSCR in a half-circle lumen phantom. (A) Tip trajectory of MSCR governed by the controlled magnetic field in the simulation and experiment, respectively. (B) The normalized distal contact force $\mathcal{F}$ plotted against normalized push distance $\mathcal{X}$. (C) The normalized magnetic field strength $\mathcal{B}$ plotted against normalized push distance $\mathcal{X}$.}
\label{fig:case1}
\end{figure}

\paragraph{Half-circle lumen phantom} 

\begin{figure}[!b]
\centering
\includegraphics[width=1\columnwidth]{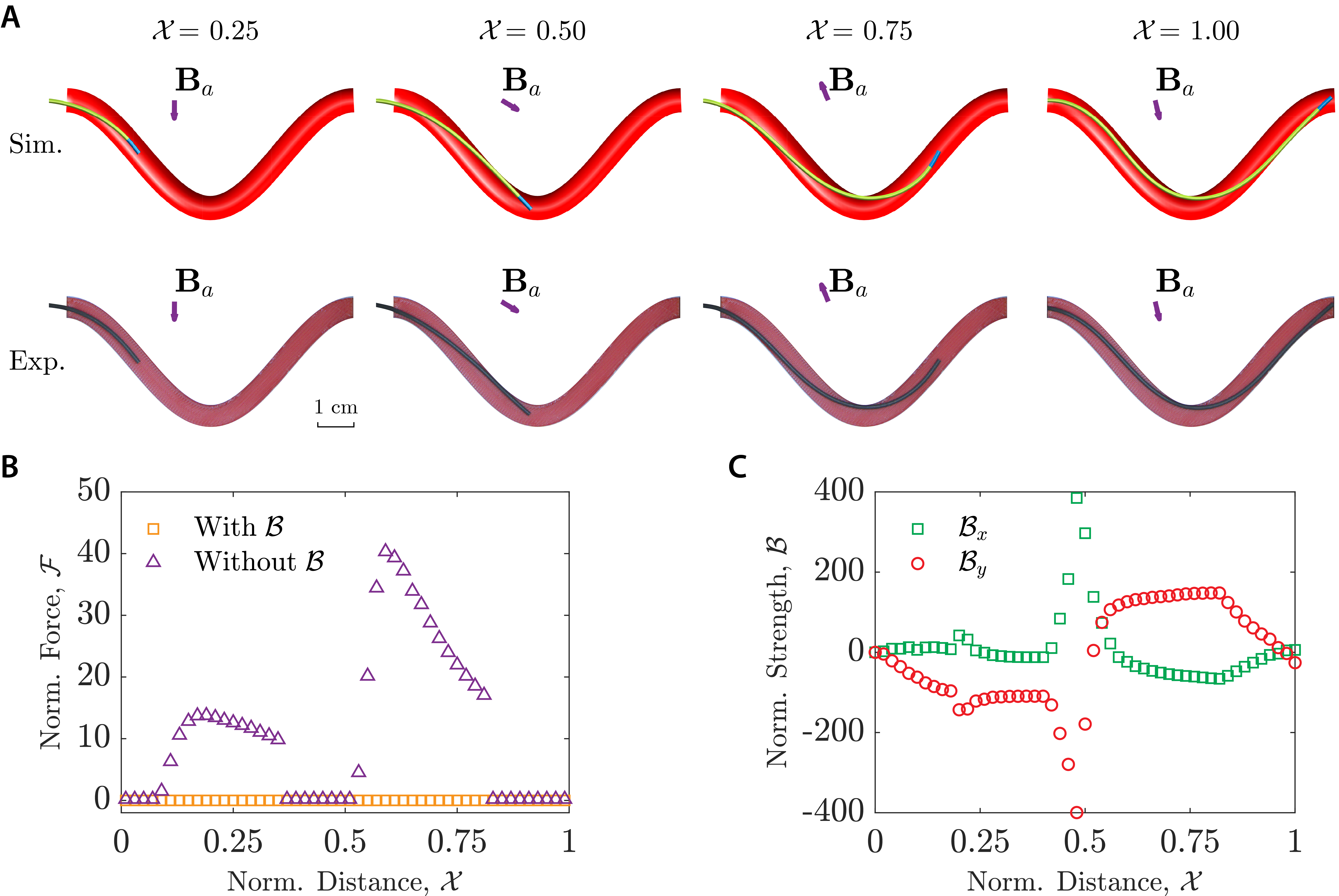}
\caption{Navigation of MSCR in a sinusoidal lumen phantom. (A) Tip trajectory of MSCR governed by the controlled magnetic field in the simulation and experiment, respectively. (B) The normalized distal contact force $\mathcal{F}$ plotted against normalized push distance $\mathcal{X}$. (C) The normalized magnetic field strength $\mathcal{B}$ plotted against normalized push distance $\mathcal{X}$.}
\label{fig:case2}
\end{figure}

We begin by evaluating the effectiveness of magnetic field control in a commonly encountered half-circle lumen phantom configuration, which mimics the aorta and was previously used for validation in the prior subsection.
Here, the lumen phantom geometry is identical to the previous passive pushing analysis, and the lumen phantom diameter is selected as $2R/L = 0.06$.
Fig. \ref{fig:case1}(A) illustrates the navigation of the MSCR with external magnetic control in comparison with the case $2$ in Fig. \ref{fig:planar_validation}, where no magnetic actuation is applied to the system.
Our model-based control method effectively regulates the tip's position and orientation, ensuring that the distal contact force $\mathcal{F}$ is eliminated, as shown in Fig.~\ref{fig:case1}(B).
The computed optimal actuation magnetic field, shown as a function of the normalized push distance $\mathcal{X}$ in Fig.\ref{fig:case1}(C), exhibits extreme nonlinearity despite the lumen phantom's constant curvature.
This nonlinearity arises from the intricate mechanics of the MSCR, including the coupling between elasticity and magnetic torques, as well as non-smooth contact interactions. These complexities make the optimal control signal highly nonlinear, further underscoring the challenges of the problem and demonstrating the robustness of the proposed active control scheme. Additionally, the computed optimal magnetic field is validated through desktop experiments (Fig.~\ref{fig:case1}(A)), where no contact between the MSCR's distal tip and the lumen phantom wall is observed under active control.
From a clinical safety perspective, the contact force applied by the MSCR's tip to lumens should typically remain as small as possible.
Our numerical framework suggests that passive pushing of MSCR poses a significant risk to patients due to the excessive tip contact forces.
This underscores the necessity of an actuation magnetic field to actively adjust the tip’s position and orientation, ensuring safer navigation.

% . Next, we implement the inverse model-based control strategy outlined in Sec.\ref{se:control}, with the results presented in Fig.~\ref{fig:case1}(B). A comparison of the two cases reveals that magnetic control enables the wire's leading end to maintain a maximum distance from the inner wall, significantly reducing contact interactions. Furthermore, at $\mathcal{X} = 0.6$, the wire under control has advanced farther along the lumen phantom than the uncontrolled wire, despite the same insertion distance. This difference arises because the controlled wire, with its leading end aligned along the centerline of the lumen phantom, experiences less resistance during traversal, resulting in a more efficient configuration.

% This effect is evident in the contact force, $\mathcal{F}$, as shown in Fig.~\ref{fig:case1}(C). Under magnetic control, the contact force remains consistently zero, unlike the uncontrolled scenario, which exhibits significant contact forces. Fig.~\ref{fig:case1}(D) shows the variation in the normalized magnetic field strength, $\mathcal{B}$, as a function of the normalized insertion distance, $\mathcal{X}$. Initially, $\mathcal{B}_x$ and $\mathcal{B}_y$ increase monotonically, in negative and positive directions, respectively, generating an external force that bends the wire to conform to the lumen phantom structure. At a critical point, $\mathcal{B}_y$ reverses direction to enable the wire’s exit from the lumen phantom, ensuring smooth traversal while minimizing contact forces.

\paragraph{Sinusoidal lumen phantom} 
Next, we examine another 2D scenario: a sinusoidal lumen phantom, as discussed before.
Again, the lumen phantom geometry is identical to the previous passive pushing analysis.
Compared with the passive pushing shown in Fig. \ref{fig:planar_validation_2}, Fig. \ref{fig:case2}(A) illustrates the tip trajectory of MSCR with the active magnetic field control. 
As shown in Fig.~\ref{fig:case2}(B), the MSCR contacts the lumen wall initially without magnetic control, resulting in relatively large contact forces. This initial contact stage dissipates when $\mathcal{X} \approx 0.5$, likely because the contact alters the tip's direction slightly (Fig.~\ref{fig:planar_validation_2}).
However, at the valley of the lumen phantom, the tangent angle changes sharply. Without external magnetic actuation to steer the MSCR, the contact forces $\mathcal{F}$ significantly increase to $40$. Such high forces could pose a substantial risk of puncturing the lumen wall, especially in vivo.
The optimal magnetic field for active control is shown in Fig.~\ref{fig:case2}(C). A pronounced change in the optimal external magnetic field $\mathbf B_a$ occurs around $\mathcal{X} \approx 0.5$, which closely aligns with the geometric features of the sinusoidal lumen phantom.
Applying this computed magnetic field effectively steers the MSCR through the lumen while maintaining a non-contact state, as validated through both numerical simulations and experimental results.

\begin{figure}[ht]
\centering
\includegraphics[width=1\columnwidth]{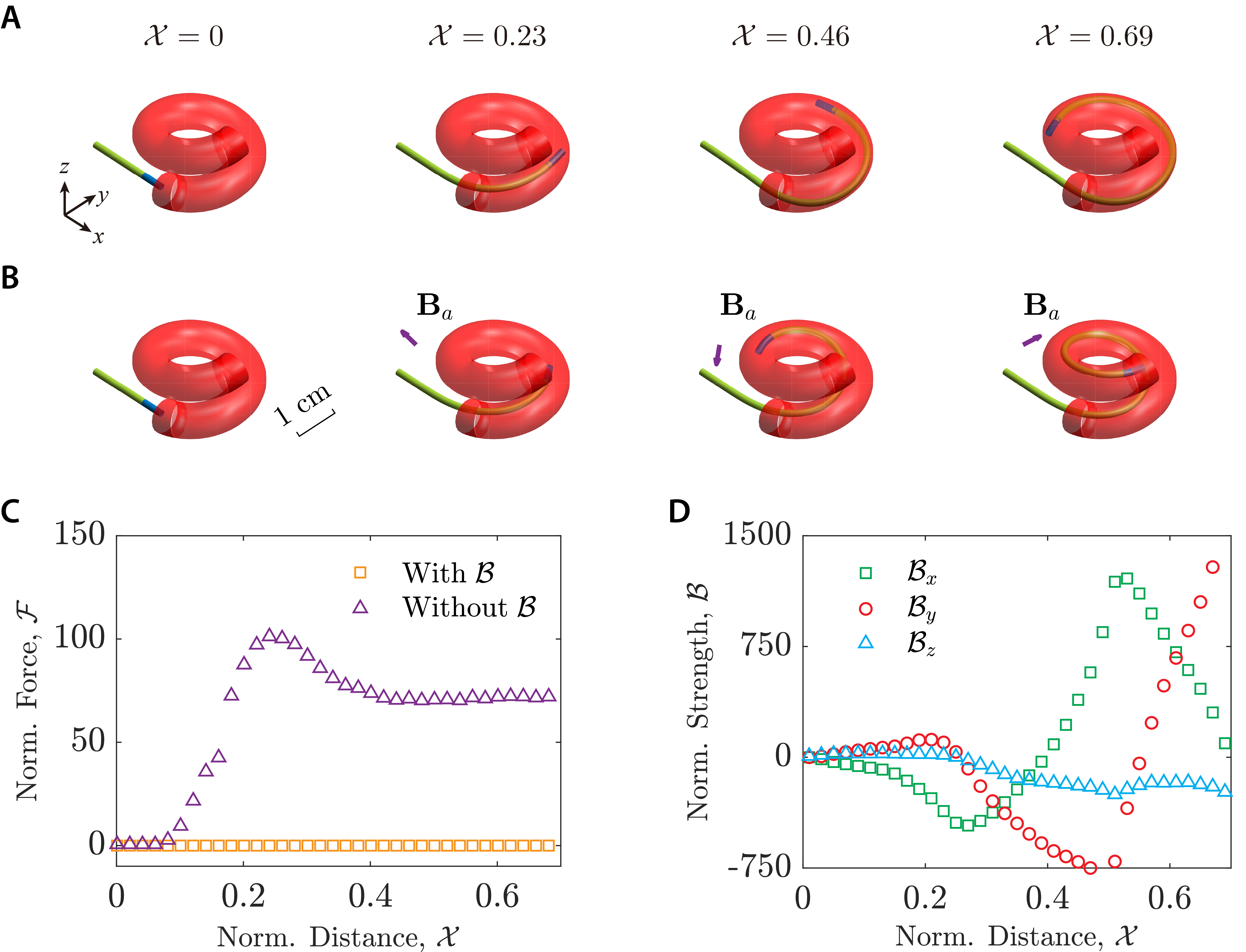}
\caption{Navigation of MSCR in a helical lumen phantom. (A) Tip trajectory of MSCR governed by the contact interaction in the absence of magnetic field control. (B) Tip trajectory of MSCR governed by the controlled magnetic field. (C) The normalized distal contact force $\mathcal{F}$ plotted against normalized push distance $\mathcal{X}$. (D) The normalized magnetic field strength $\mathcal{B}$ plotted against normalized push distance $\mathcal{X}$. }
\label{fig:case3}
\end{figure}

\subsection{Active Control of the MSCR Navigating in 3D lumen phantoms}

In this part, we apply the proposed model-based control algorithm to compute the actuation magnetic field required to navigate the MSCR through 3D tubular lumens, providing a more generalized scenario to demonstrate the effectiveness of our approach. Due to experimental limitations, only numerical demonstrations are presented. We investigate the navigation of MSCR in helical lumen phantoms, which commonly appear in coronary arteries, and explore its steerability through bifurcated lumen phantoms to assess control performance in complex vascular environments.

\paragraph{Helical lumen phantom} 
The helical lumen phantom is characterized by a radius of  $\hat{R}_{h}/L = 0.1$, a pitch of $ \hat{\lambda}_{h}/L = 0.1 $ (thus $\hat{P}_{h} = \hat{\lambda}_{h} / 2\pi$), and total length of $\hat{S}_{h} / L = 0.9$, resulting in
\begin{equation}
\begin{cases}
  {\hat{X}_{h}}  / L&= {\hat{R}_{h}} \cos (\omega - \frac {\pi} {2}) / L, \\
  {\hat{Y}_{h}}  / L&=  {\hat{R}_{h}} \sin (\omega - \frac {\pi} {2}) / L +  {\hat{R}_{h}} / L,\\
{\hat{Z}_{h}} / L &=  {\hat{P}_{h}}  {\omega }/L,
 \end{cases}
 \; \; \; \mathrm{with} \; \omega \in [0, \frac { \hat{S}_{h}} {\sqrt{ \hat{R}_{h}^2 + {\hat{P}_{h}^2}  }}],
\end{equation}
and the lumen phantom diameter is $2R/L = 0.06$.
Figs.~\ref{fig:case3}(A) and (B) illustrate the tip trajectories with and without magnetic field control, respectively. As shown in Fig.~\ref{fig:case3}(C), the maximum normalized contact force $\mathcal{F}$ is significantly higher than in planar cases, highlighting the increased risk of the MSCR’s tip puncturing the lumen wall due to the complex geometry of the lumen phantom. However, with the computed actuation magnetic field shown in Fig.~\ref{fig:case3}(D), our active control method effectively prevents contact between the MSCR’s distal tip and the lumen phantom wall, thereby eliminating the contact force $\mathcal{F}$.

\begin{figure}[!h]
\centering
\includegraphics[width=1.0\columnwidth]{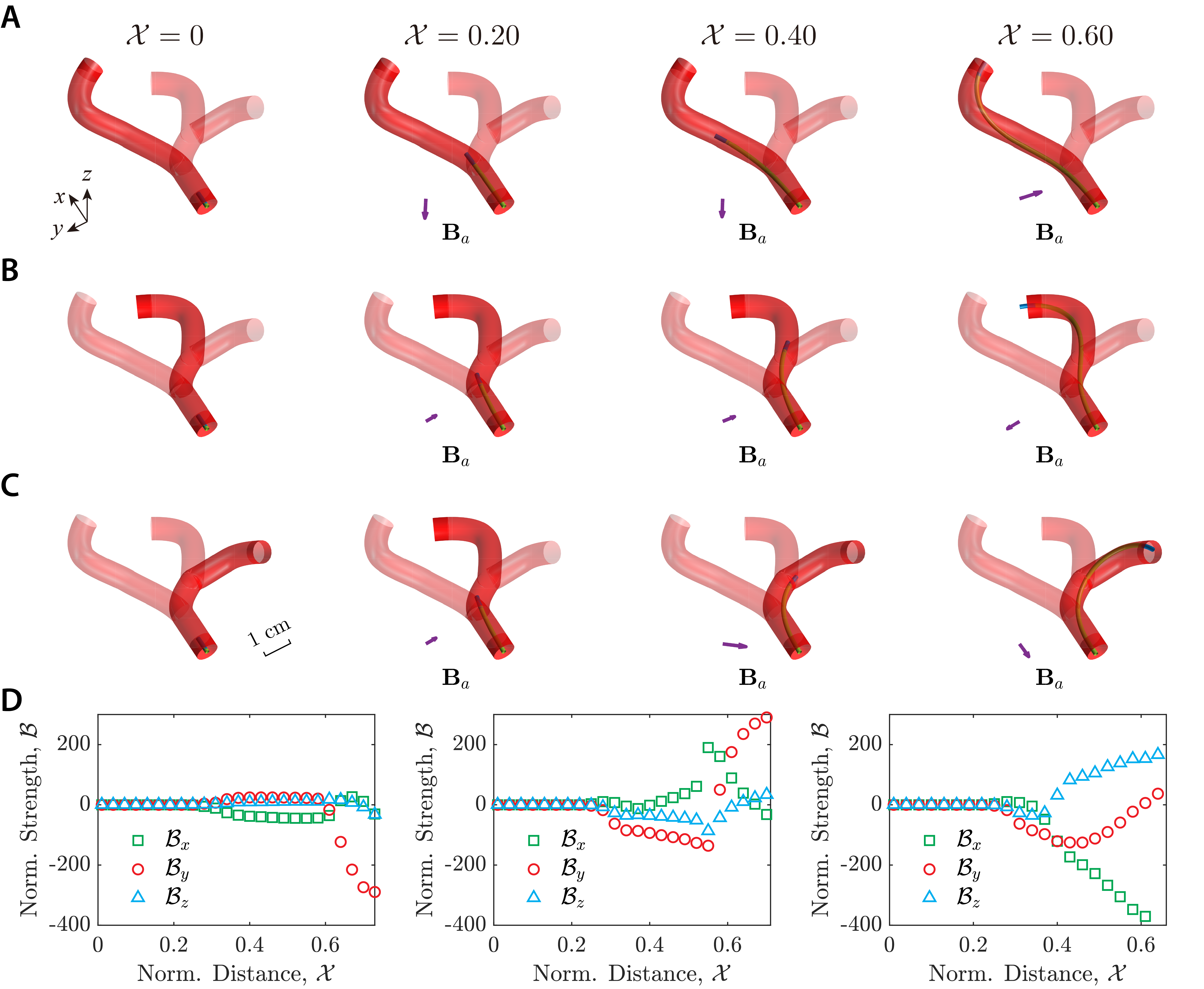}
\caption{Magnetic navigation of MSCR in the multi-bifurcated lumen phantoms. (A) Left branch. (B) Middle branch. (C) Right branch. (D) The normalized magnetic field strength $\mathcal{B}$ plotted against normalized push distance $\mathcal{X}$ for three branches.}
\label{fig:bifurcation}
\end{figure}

\paragraph{Multi-bifurcated lumen phantoms} 
The model-based control method not only navigates the MSCR through curved lumen phantoms but also enables it to select the correct branch when encountering multi-bifurcated junctions, which are common in the human circulatory system.
Here, we present a 3D lumen phantom with three branches, as shown in Fig. \ref{fig:bifurcation}, and the lumen phantom diameter is $2R/L = 0.06$.
Using the proposed model-based control method, we compute the optimal magnetic field to steer the MSCR's tip along one of the three branches, left, middle, or right, based on specific requirements. This demonstrates the versatility of the control strategy in navigating MSCR within complex yet realistic in-vivo environments.

\subsection{Computational speed analysis}
The accuracy of the proposed numerical framework and the model-based control method have been comprehensively studied in this manuscript. Here, We conduct a full analysis on the computational efficiency of the proposed methodology.

\begin{figure}[!h]
\centering
\includegraphics[width=1\columnwidth]{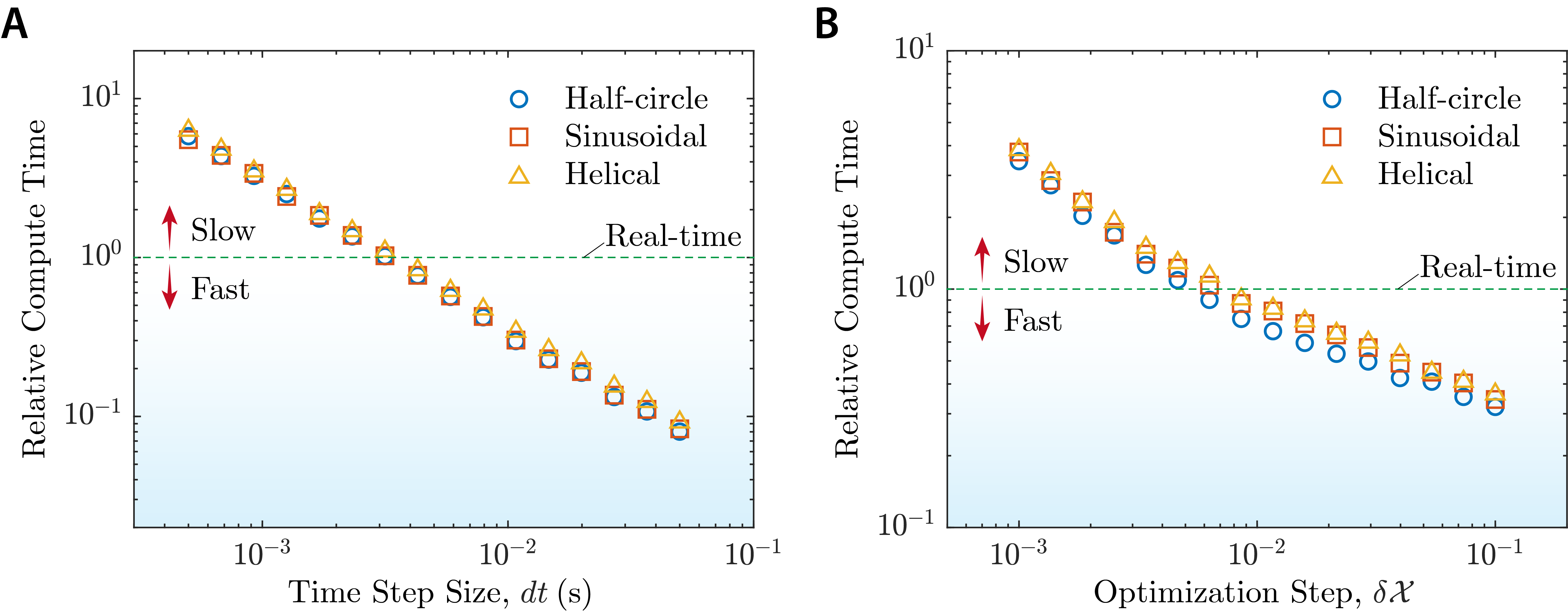}
\caption{(A) Relative compute time, which is the ratio between the computational time and the simulation clock time, as a function of time step size. (B) Relative compute time as a function of optimization step.}
\label{fig:computation}
\end{figure}

First, we evaluate the proposed reduced-order framework for simulating the passive pushing of the MSCR. As shown in Fig.~\ref{fig:computation} A, our simulation framework can support large time steps, e.g., $dt \le 0.1$ s for all cases studied in this manuscript. This benefits from its fully implicit numerical formulation. Notably, when $dt > 0.003$ s, the relative compute time, which is defined as the ratio between the computational time and the simulation clock time, falls below $1.0$. This indicates that our simulation not only achieves but also surpasses real-time performance. The computational efficiency of the numerical framework lays the foundation for the real-time performance of the proposed model-based control methodology. Next, in Fig.~\ref{fig:computation} B, we study the computational efficiency of the proposed active control method. Here, the optimization step, $\delta \mathcal{X}$, represents the normalized displacement of the tip in the lumen phantom between two consecutive updates of the actuation magnetic field. Here, the normalized total length of MSCR is $\mathcal{X}=1.0$. In all three cases, we find that our model-based control method can effectively eliminate contact if $\delta \mathcal{X} \le 0.1$. In addition, when $\delta \mathcal{X}$ is larger than $0.005$, the model-based control can achieve real-time performance. These findings demonstrate that our approach strikes an optimal balance between computational efficiency and accuracy, enabling real-time MSCR applications in clinical intervention procedures.
 
\section{Conclusions} \label{sec:discussion}

In this paper, we present a model-based control framework for navigating MSCR in confined lumens.
Building on hard-magnetic elastic rod theory, we developed a reduced-order simulation framework that accurately captures the geometrically nonlinear magneto-elastic deformation and boundary nonlinear contact interactions for MSCR navigating through lumen phantoms.
Next, we effectively integrate control constraints into the mechanics model, enabling real-time model-based control while leveraging the computational efficiency of our reduced-order numerical simulations.
Our proposed control framework computes the required actuation magnetic field to regulate the MSCR's distal tip, ensuring contact-free navigation within the lumen phantom.
This approach significantly mitigates the risks associated with MSCR applications in intervention procedures, enhancing both safety and reliability.
The proposed model-based control method was rigorously validated through numerical simulations and experimental studies, demonstrating its effectiveness.
Our methodology successfully addresses the challenge of utilizing mechanics models for optimal control of MSCRs.

For future work, we plan to integrate the proposed methodology into robotic systems to enable full automation of MSCR applications in clinical intervention procedures.
Additionally, our simplified penalty-based energy formulation for solving optimal control problems with mechanics models highlights a promising direction for incorporating the predictive power of classical mechanics into controller design, paving the way for automated solutions in soft matter manipulation.

\section*{Acknowledgments}

L. Wang acknowledges the support from the National Natural Science Foundation of China (Grant No. 12388101, 12272369).  M. Liu acknowledges the start-up funding from The University of Birmingham, UK. W. Huang acknowledges the start-up funding from Newcastle University, UK.

\appendix

\section{Video}
\label{sec:AppendixA}

We provide a video as supplementary material to illustrate our result.

%\newpage % separate Appendix and references
\bibliographystyle{elsarticle-harv}
\bibliography{paper}

\begin{thebibliography}{50}
\expandafter\ifx\csname natexlab\endcsname\relax\def\natexlab#1{#1}\fi
\expandafter\ifx\csname url\endcsname\relax
  \def\url#1{\texttt{#1}}\fi
\expandafter\ifx\csname urlprefix\endcsname\relax\def\urlprefix{URL }\fi

\bibitem[{Appireddy et~al.(2016)Appireddy, Zerna, Menon, and
  Goyal}]{appireddy2016endovascular}
Appireddy, R., Zerna, C., Menon, B.~K., Goyal, M., 2016. Endovascular
  interventions in acute ischemic stroke: recent evidence, current challenges,
  and future prospects. Current atherosclerosis reports 18, 1--11.

\bibitem[{Audoly and Pomeau(2000)}]{audoly2000elasticity}
Audoly, B., Pomeau, Y., 2000. Elasticity and geometry. In: Peyresq Lectures on
  Nonlinear Phenomena. World Scientific, pp. 1--35.

\bibitem[{Bergou et~al.(2010)Bergou, Audoly, Vouga, Wardetzky, and
  Grinspun}]{bergou2010discrete}
Bergou, M., Audoly, B., Vouga, E., Wardetzky, M., Grinspun, E., 2010. Discrete
  viscous threads. ACM Transactions on graphics (TOG) 29~(4), 1--10.

\bibitem[{Bergou et~al.(2008)Bergou, Wardetzky, Robinson, Audoly, and
  Grinspun}]{bergou2008discrete}
Bergou, M., Wardetzky, M., Robinson, S., Audoly, B., Grinspun, E., 2008.
  Discrete elastic rods. In: ACM SIGGRAPH 2008 papers. pp. 1--12.

\bibitem[{Chen and Wang(2020)}]{chen2020theoretical}
Chen, W., Wang, L., 2020. Theoretical modeling and exact solution for extreme
  bending deformation of hard-magnetic soft beams. Journal of Applied Mechanics
  87~(4), 041002.

\bibitem[{Chen et~al.(2021)Chen, Wang, Yan, and Luo}]{chen2021three}
Chen, W., Wang, L., Yan, Z., Luo, B., 2021. Three-dimensional large-deformation
  model of hard-magnetic soft beams. Composite Structures 266, 113822.

\bibitem[{Chen et~al.(2020{\natexlab{a}})Chen, Yan, and Wang}]{chen2020complex}
Chen, W., Yan, Z., Wang, L., 2020{\natexlab{a}}. Complex transformations of
  hard-magnetic soft beams by designing residual magnetic flux density. Soft
  Matter 16~(27), 6379--6388.

\bibitem[{Chen et~al.(2020{\natexlab{b}})Chen, Yan, and
  Wang}]{chen2020mechanics}
Chen, W., Yan, Z., Wang, L., 2020{\natexlab{b}}. On mechanics of functionally
  graded hard-magnetic soft beams. International Journal of Engineering Science
  157, 103391.

\bibitem[{Choi et~al.(2021)Choi, Tong, Jawed, and Joo}]{choi2021implicit}
Choi, A., Tong, D., Jawed, M.~K., Joo, J., 2021. Implicit contact model for
  discrete elastic rods in knot tying. Journal of Applied Mechanics 88~(5),
  051010.

\bibitem[{Dadgar-Rad et~al.(2024)Dadgar-Rad, Hemmati, and
  Hossain}]{dadgar2024three}
Dadgar-Rad, F., Hemmati, A., Hossain, M., 2024. A three-dimensional micropolar
  beam model with application to the finite deformation analysis of
  hard-magnetic soft beams. International Journal of Solids and Structures 290,
  112662.

\bibitem[{Dadgar-Rad and Hossain(2022)}]{dadgar2022finite}
Dadgar-Rad, F., Hossain, M., 2022. Finite deformation analysis of hard-magnetic
  soft materials based on micropolar continuum theory. International Journal of
  Solids and Structures 251, 111747.

\bibitem[{Duan et~al.(2023)Duan, Akinyemi, Du, Ma, Chen, Wang, Omisore, Luo,
  Wang, and Wang}]{duan2023technical}
Duan, W., Akinyemi, T., Du, W., Ma, J., Chen, X., Wang, F., Omisore, O., Luo,
  J., Wang, H., Wang, L., 2023. Technical and clinical progress on
  robot-assisted endovascular interventions: A review. Micromachines 14~(1),
  197.

\bibitem[{Garcia-Gonzalez and Hossain(2021)}]{garcia2021microstructural}
Garcia-Gonzalez, D., Hossain, M., 2021. Microstructural modelling of
  hard-magnetic soft materials: Dipole--dipole interactions versus zeeman
  effect. Extreme Mechanics Letters 48, 101382.

\bibitem[{Goyal et~al.(2016)Goyal, Yu, Menon, Dippel, Hacke, Davis, Fisher,
  Yavagal, Turjman, Ross, et~al.}]{goyal2016endovascular}
Goyal, M., Yu, A.~Y., Menon, B.~K., Dippel, D.~W., Hacke, W., Davis, S.~M.,
  Fisher, M., Yavagal, D.~R., Turjman, F., Ross, J., et~al., 2016. Endovascular
  therapy in acute ischemic stroke: challenges and transition from trials to
  bedside. Stroke 47~(2), 548--553.

\bibitem[{Gunduz et~al.(2021)Gunduz, Albadawi, and Oklu}]{gunduz2021robotic}
Gunduz, S., Albadawi, H., Oklu, R., 2021. Robotic devices for minimally
  invasive endovascular interventions: a new dawn for interventional radiology.
  Advanced Intelligent Systems 3~(2), 2000181.

\bibitem[{Huang et~al.(2023{\natexlab{a}})Huang, Liu, and
  Hsia}]{huang2023modeling}
Huang, W., Liu, M., Hsia, K., 2023{\natexlab{a}}. Modeling of magnetic cilia
  carpet robots using discrete differential geometry formulation. Extreme
  Mechanics Letters 59, 101967.

\bibitem[{Huang et~al.(2023{\natexlab{b}})Huang, Liu, and
  Hsia}]{huang2023discrete}
Huang, W., Liu, M., Hsia, K.~J., 2023{\natexlab{b}}. A discrete model for the
  geometrically nonlinear mechanics of hard-magnetic slender structures.
  Extreme Mechanics Letters 59, 101977.

\bibitem[{Huang et~al.(2024)Huang, Xu, and Liu}]{huang2024dynamic}
Huang, W., Xu, P., Liu, Z., 2024. Dynamic modeling of a sliding ring on an
  elastic rod with incremental potential formulation. Journal of Applied
  Mechanics, 1--13.

\bibitem[{Hwang et~al.(2020)Hwang, Kim, and Choi}]{hwang2020review}
Hwang, J., Kim, J.-y., Choi, H., 2020. A review of magnetic actuation systems
  and magnetically actuated guidewire-and catheter-based microrobots for
  vascular interventions. Intelligent Service Robotics 13, 1--14.

\bibitem[{Jawed et~al.(2018)Jawed, Novelia, and O'Reilly}]{jawed2018primer}
Jawed, M.~K., Novelia, A., O'Reilly, O.~M., 2018. A primer on the kinematics of
  discrete elastic rods. Springer.

\bibitem[{Khoshnam et~al.(2015)Khoshnam, Skanes, and
  Patel}]{khoshnam2015modeling}
Khoshnam, M., Skanes, A.~C., Patel, R.~V., 2015. Modeling and estimation of tip
  contact force for steerable ablation catheters. IEEE Transactions on
  Biomedical Engineering 62~(5), 1404--1415.

\bibitem[{Kim et~al.(2022)Kim, Genevriere, Harker, Choe, Balicki, Regenhardt,
  Vranic, Dmytriw, Patel, and Zhao}]{kim2022telerobotic}
Kim, Y., Genevriere, E., Harker, P., Choe, J., Balicki, M., Regenhardt, R.~W.,
  Vranic, J.~E., Dmytriw, A.~A., Patel, A.~B., Zhao, X., 2022. Telerobotic
  neurovascular interventions with magnetic manipulation. Science Robotics
  7~(65), eabg9907.

\bibitem[{Kim et~al.(2019)Kim, Parada, Liu, and Zhao}]{kim2019ferromagnetic}
Kim, Y., Parada, G.~A., Liu, S., Zhao, X., 2019. Ferromagnetic soft continuum
  robots. Science robotics 4~(33), eaax7329.

\bibitem[{Kim and Zhao(2022)}]{kim2022magnetic}
Kim, Y., Zhao, X., 2022. Magnetic soft materials and robots. Chemical reviews
  122~(5), 5317--5364.

\bibitem[{Li et~al.(2024{\natexlab{a}})Li, Chen, and Wang}]{li2024model}
Li, J., Chen, H., Wang, L., 2024{\natexlab{a}}. Model-guided navigation of
  magnetic soft guidewire for safe endovascular surgery. Journal of the
  Mechanics and Physics of Solids, 105731.

\bibitem[{Li and Wang(2024)}]{li2024modeling}
Li, J., Wang, L., 2024. Modeling magnetic soft continuum robot in nonuniform
  magnetic fields via energy minimization. International Journal of Mechanical
  Sciences 282, 109688.

\bibitem[{Li et~al.(2024{\natexlab{b}})Li, Feng, Zhang, Fang, Zhang, and
  Liang}]{li2024modeling1}
Li, P., Feng, J., Zhang, X., Fang, D., Zhang, J., Liang, C.,
  2024{\natexlab{b}}. Modeling and experimental study of the intervention
  forces between the guidewire and blood vessels. Medical Engineering \&
  Physics 127, 104166.

\bibitem[{Li et~al.(2023)Li, Yu, Liu, Zhu, Wang, Sun, Liu, and
  Yuan}]{li2023mechanics}
Li, X., Yu, W., Liu, J., Zhu, X., Wang, H., Sun, X., Liu, J., Yuan, H., 2023. A
  mechanics model of hard-magnetic soft rod with deformable cross-section under
  three-dimensional large deformation. International Journal of Solids and
  Structures 279, 112344.

\bibitem[{Liu et~al.(2023)Liu, Yang, Li, and Xu}]{liu2023meshfree}
Liu, J., Yang, Y., Li, M., Xu, F., 2023. A meshfree model of hard-magnetic soft
  materials. International Journal of Mechanical Sciences 258, 108566.

\bibitem[{Martin et~al.(2020)Martin, Scaglioni, Norton, Subramanian, Arezzo,
  Obstein, and Valdastri}]{martin2020enabling}
Martin, J.~W., Scaglioni, B., Norton, J.~C., Subramanian, V., Arezzo, A.,
  Obstein, K.~L., Valdastri, P., 2020. Enabling the future of colonoscopy with
  intelligent and autonomous magnetic manipulation. Nature machine intelligence
  2~(10), 595--606.

\bibitem[{Moezi et~al.(2024)Moezi, Sedaghati, and
  Rakheja}]{moezi2024development}
Moezi, S.~A., Sedaghati, R., Rakheja, S., 2024. Development of a novel
  nonlinear model and control strategy for soft continuum robots featuring hard
  magnetoactive elastomers. Smart Materials and Structures 33~(4), 045025.

\bibitem[{Muller et~al.(1992)Muller, Shamir, Ellis, and
  Topol}]{muller1992peripheral}
Muller, D.~W., Shamir, K.~J., Ellis, S.~G., Topol, E.~J., 1992. Peripheral
  vascular complications after conventional and complex percutaneous coronary
  interventional procedures. The American journal of cardiology 69~(1), 63--68.

\bibitem[{Munich et~al.(2019)Munich, Vakharia, and Levy}]{munich2019overview}
Munich, S.~A., Vakharia, K., Levy, E.~I., 2019. Overview of mechanical
  thrombectomy techniques. Neurosurgery 85~(suppl\_1), S60--S67.

\bibitem[{Ruszczynski(2011)}]{ruszczynski2011nonlinear}
Ruszczynski, A., 2011. Nonlinear optimization. Princeton university press.

\bibitem[{Sano et~al.(2022)Sano, Pezzulla, and Reis}]{sano2022kirchhoff}
Sano, T.~G., Pezzulla, M., Reis, P.~M., 2022. A kirchhoff-like theory for hard
  magnetic rods under geometrically nonlinear deformation in three dimensions.
  Journal of the Mechanics and Physics of Solids 160, 104739.

\bibitem[{Stewart and Anand(2023)}]{stewart2023magneto}
Stewart, E.~M., Anand, L., 2023. Magneto-viscoelasticity of hard-magnetic
  soft-elastomers: Application to modeling the dynamic snap-through behavior of
  a bistable arch. Journal of the Mechanics and Physics of Solids 179, 105366.

\bibitem[{Tong et~al.(2023)Tong, Choi, Joo, and Jawed}]{tong2023fully}
Tong, D., Choi, A., Joo, J., Jawed, M.~K., 2023. A fully implicit method for
  robust frictional contact handling in elastic rods. Extreme Mechanics Letters
  58, 101924.

\bibitem[{Wang et~al.(2022)Wang, Guo, and Zhao}]{wang2022magnetic}
Wang, L., Guo, C.~F., Zhao, X., 2022. Magnetic soft continuum robots with
  contact forces. Extreme Mechanics Letters 51, 101604.

\bibitem[{Wang et~al.(2020)Wang, Kim, Guo, and Zhao}]{wang2020hard}
Wang, L., Kim, Y., Guo, C.~F., Zhao, X., 2020. Hard-magnetic elastica. Journal
  of the Mechanics and Physics of Solids 142, 104045.

\bibitem[{Wang et~al.(2021)Wang, Zheng, Harker, Patel, Guo, and
  Zhao}]{wang2021evolutionary}
Wang, L., Zheng, D., Harker, P., Patel, A.~B., Guo, C.~F., Zhao, X., 2021.
  Evolutionary design of magnetic soft continuum robots. Proceedings of the
  National Academy of Sciences 118~(21), e2021922118.

\bibitem[{Wang et~al.(2015)Wang, Chen, Tai, Xu, and Shih}]{wang2015study}
Wang, Y., Chen, R.~K., Tai, B.~L., Xu, K., Shih, A.~J., 2015. Study of
  insertion force and deformation for suturing with precurved niti guidewire.
  Journal of Biomechanical Engineering 137~(4), 041004.

\bibitem[{Wang et~al.(2024)Wang, Qin, Luo, Tian, and Hu}]{wang2024dynamic}
Wang, Y., Qin, Y., Luo, K., Tian, Q., Hu, H., 2024. Dynamic modeling and
  simulation of hard-magnetic soft beams interacting with environment via
  high-order finite elements of ancf. International Journal of Engineering
  Science 202, 104102.

\bibitem[{Wu et~al.(2020)Wu, Hu, Ze, Sitti, and Zhao}]{wu2020multifunctional}
Wu, S., Hu, W., Ze, Q., Sitti, M., Zhao, R., 2020. Multifunctional magnetic
  soft composites: A review. Multifunctional materials 3~(4), 042003.

\bibitem[{Yan et~al.(2022)Yan, Abbasi, and Reis}]{yan2022comprehensive}
Yan, D., Abbasi, A., Reis, P.~M., 2022. A comprehensive framework for
  hard-magnetic beams: reduced-order theory, 3d simulations, and experiments.
  International Journal of Solids and Structures 257, 111319.

\bibitem[{Yang et~al.(2023{\natexlab{a}})Yang, Zhou, Zhao, Huang, Wang, Hsia,
  and Liu}]{yang2023morphing}
Yang, X., Zhou, Y., Zhao, H., Huang, W., Wang, Y., Hsia, K.~J., Liu, M.,
  2023{\natexlab{a}}. Morphing matter: From mechanical principles to robotic
  applications. Soft Sci 3~(4), 38.

\bibitem[{Yang et~al.(2023{\natexlab{b}})Yang, Yang, Cao, Cui, and
  Zhang}]{yang2023magnetically}
Yang, Z., Yang, H., Cao, Y., Cui, Y., Zhang, L., 2023{\natexlab{b}}.
  Magnetically actuated continuum medical robots: A review. Advanced
  intelligent systems 5~(6), 2200416.

\bibitem[{Yao et~al.(2023)Yao, Cao, Ju, Sun, Liu, Han, and
  Li}]{yao2023adaptive}
Yao, J., Cao, Q., Ju, Y., Sun, Y., Liu, R., Han, X., Li, L., 2023. Adaptive
  actuation of magnetic soft robots using deep reinforcement learning. Advanced
  Intelligent Systems 5~(2), 2200339.

\bibitem[{Ye et~al.(2021)Ye, Li, and Zhang}]{ye2021magttice}
Ye, H., Li, Y., Zhang, T., 2021. Magttice: A lattice model for hard-magnetic
  soft materials. Soft Matter 17~(13), 3560--3568.

\bibitem[{Zhao et~al.(2019)Zhao, Kim, Chester, Sharma, and
  Zhao}]{zhao2019mechanics}
Zhao, R., Kim, Y., Chester, S.~A., Sharma, P., Zhao, X., 2019. Mechanics of
  hard-magnetic soft materials. Journal of the Mechanics and Physics of Solids
  124, 244--263.

\bibitem[{Zhao et~al.(2022)Zhao, Mei, Luo, Mao, Zhao, Liu, and
  Wu}]{zhao2022remote}
Zhao, Y., Mei, Z., Luo, X., Mao, J., Zhao, Q., Liu, G., Wu, D., 2022. Remote
  vascular interventional surgery robotics: A literature review. Quantitative
  Imaging in Medicine and Surgery 12~(4), 2552.

\end{thebibliography}

\end{document}